\documentclass[preprint,10pt]{elsarticle}

\usepackage{amsmath,amssymb,bm,amsthm}
\usepackage{subfigure}
\usepackage{graphicx}
\usepackage{dcolumn}
\usepackage{bm}
\usepackage[mathlines]{lineno}
\usepackage{booktabs}
\usepackage{color}
\newcounter{one}
\usepackage{url}

\usepackage{textcase}

\usepackage{colortbl}
\usepackage{tabularx}
\usepackage{verbatim}
\usepackage{multirow}

\usepackage[T1]{fontenc} 
\usepackage{lmodern}
\usepackage{bbm}
\usepackage[utf8]{inputenc}
\usepackage{amsfonts}
\usepackage{array}

\usepackage{bbm}
\usepackage{enumitem}
\usepackage{umoline}
\usepackage[usenames,svgnames]{xcolor}
\usepackage{natbib}
\usepackage{hyperref}
\hypersetup{
     colorlinks=true,       		
     linkcolor=Navy,          	
     citecolor=Navy,            
     filecolor=Navy,      		
     urlcolor=Navy,           	
    runcolor=cyan,
 }

\usepackage{graphicx}
\usepackage{amsfonts}
\usepackage{amssymb}
\usepackage{amsmath}
\usepackage{bbm}
\usepackage{enumitem}

\newcommand{\bra}[1]{\langle #1 |}
\newcommand{\ket}[1]{| #1 \rangle}

\newcommand{\tr}[0]{ {\rm tr}}

\newcommand{\half}[1]{{ \rm  h}}
\newcommand{\Oorderof}{\mathcal{O}}
\newcommand{\orderof}[1]{\Oorderof(#1)} 

\newcommand{\for}[0]{\quad  {\rm for} \quad}

\newcommand{\dist}{{\rm dist}}

\newcommand{\Cl}{\mathcal{C}}
\newcommand{\Gc}{\mathcal{G}}

\newcommand{\ban}{{\mathcal{B}}}

\newcommand{\Der}{{\mathcal{D}}}

\newcommand{\Fp}{\mathcal{F}}

\newcommand{\ed}{X}

\def\beq{\begin{equation}}
\def\eeq{\end{equation}}
\def\nbeq{\begin{equation*}}
\def\neeq{\end{equation*}}
\def\<{\langle}
\def\>{\rangle}

\def\tr{{\rm tr}}

\theoremstyle{theorem}
\newtheorem{theorem}{Theorem}
\newtheorem{lemma}{Lemma}
\newtheorem{corol}{Corollary}
\newtheorem{definition}{Definition}  
\newtheorem{prop}{Proposition}  

\usepackage{ulem}

\usepackage{mathtools}
\def\multiset#1#2{\ensuremath{\left(\kern-.3em\left(\genfrac{}{}{0pt}{}{#1}{#2}\right)\kern-.3em\right)}}

\newcommand{\copyave}[2]{\left \langle #2 \right \rangle_0^{#1}}

\usepackage{amssymb}

\journal{Annals of Physics}

\begin{document}

\begin{frontmatter}



\title{Gaussian concentration bound and Ensemble equivalence in generic quantum many-body systems including long-range interactions}


\author{Tomotaka Kuwahara}
\address{Mathematical Science Team, RIKEN Center for Advanced Intelligence Project (AIP),1-4-1 Nihonbashi, Chuo-ku, Tokyo 103-0027, Japan}
\address{interdisciplinary Theoretical \& Mathematical Sciences Program (iTHEMS) RIKEN 2-1, Hirosawa, Wako, Saitama 351-0198, Japan}

\author{Keiji Saito}
\address{Department of Physics, Keio University, 3-14-1 Hiyoshi, Kohoku-ku, Yokohama, Japan 223-8522}

\begin{abstract}

This work explores fundamental statistical and thermodynamic properties of short-and long-range-interacting systems.
The purpose of this study is twofold. 
Firstly, we rigorously prove that the probability distribution of arbitrary few-body observables is restricted by a Gaussian concentration bound (or Chernoff--Hoeffding inequality) above some threshold temperature.
This bound is then derived for arbitrary Gibbs states of systems that include long-range interactions
Secondly, we establish a quantitative relationship between the concentration bound of the Gibbs state and the equivalence of canonical and micro-canonical ensembles.
We then evaluate the difference in the averages of thermodynamic properties between the canonical and the micro-canonical ensembles.
Under the assumption of the Gaussian concentration bound on the canonical ensemble, the difference between the ensemble descriptions is upper-bounded by $\left[n^{-1} \log (n^{3/2}\Delta^{-1})\right]^{1/2}$ with $n$ being the system size and $\Delta$ being the width of the energy shell of the micro-canonical ensemble
This limit gives a non-trivial upper bound \textit{exponentially small energy width} with respect to the system size.  
By combining these two results, we prove the ensemble equivalence as well as the weak eigenstate thermalization in arbitrary long-range-interacting systems above a threshold temperature.

\end{abstract}

\begin{keyword}

Long-range-interacting systems, Concentration bound, Chernoff--Hoeffding inequality, 
Ensemble equivalence, Eigenstate thermalization hypothesis, Weak eigenstate thermalization   

{~}\\

 \PACS 
05.30.Ch, \
05.30.-d, \
65.40.Gr, \
02.50.-r, 

{~}\\

Highlights: 


* Foundational treatment of thermodynamics and statistical mechanics in generic long-range-interacting systems

* Proof of a Gaussian concentration bound on the probability distribution of observables

* Equivalence between canonical and micro-canonical ensembles proven for long-range-interacting systems

* Weak eigenstate thermalization in long-range-interacting systems is proven


* The width of the energy shell can be taken exponentially small with respect to the system size




\end{keyword}

\end{frontmatter}



\section{Introduction} \label{sec:introduction}

In recent years, systems that include long-range interactions have become ubiquitous in various experimental setups for studying atomic, molecular, and optical systems~\cite{richerme2014non,
RevModPhys.80.885,
britton2012engineered,
doi:10.1080/00018730701223200,
RevModPhys.82.2313,
yan2013observation,
bendkowsky2009observation}.
These systems often exhibit novel physics that do not appear in short-range interacting systems~\cite{PhysRevLett.106.130601,PhysRevLett.111.260401,PhysRevLett.110.185302,PhysRevLett.119.023001,Neyenhuise1700672}.
In both experimental and theoretical contexts, long-range-interacting systems play crucial roles in modern physics.
Most existing analyses of short-range interacting systems require non-trivial modifications before they can be applied to systems with long-range interactions.

\begin{figure}[tt]
\centering
{
\includegraphics[clip, scale=0.4]{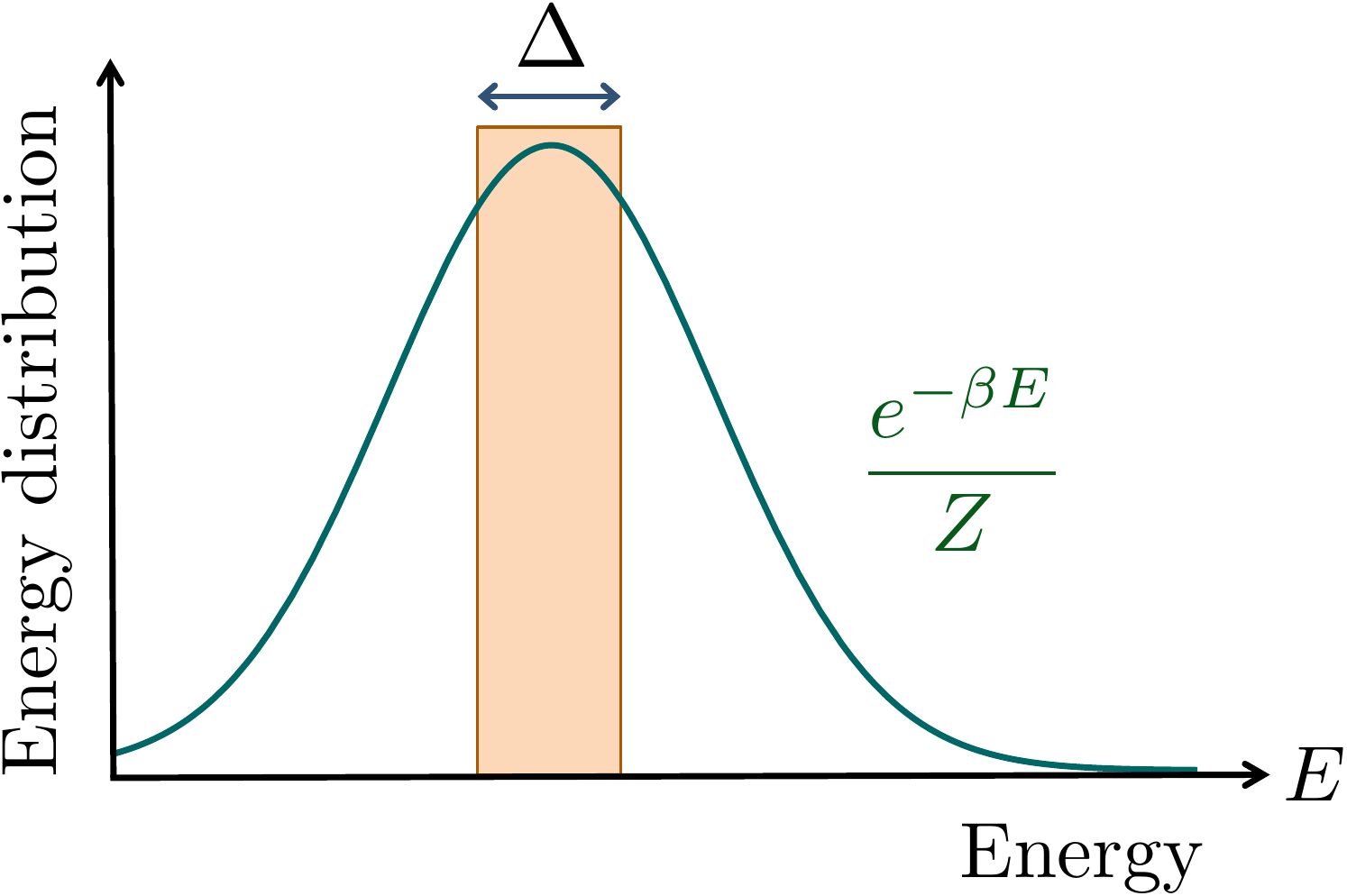}
}
\caption{
Schematic of a canonical and a micro-canonical ensemble. The canonical ensemble is characterized by the distribution $e^{-\beta E}/Z$ with $Z$ being the partition function (green curve). 
The micro-canonical ensemble is defined by the uniform distribution within an energy shell (orange region). 
The precise definitions are shown in Eq.~\eqref{def:cano_micro}. 
The problem of ensemble equivalence is about whether these two ensembles have similar expectation values for thermodynamic quantities such as magnetization.    
Our goal is to quantitatively evaluate the dependence of ensemble equivalence on the system size $n$ and the width of the energy shell $\Delta$. 
Theorem~\ref{main_theorem_Ensemble Equivalence} shows that the ensemble equivalence of Gibbs states is deeply related to the concentration bound~\eqref{ineq:Chernoff_bound}.
By proving the concentration bound with $\gamma=2$ in generic long-range-interacting systems above a threshold temperature (Corollary~\ref{main_corol_CH_inequality}), we rigorously prove the equivalence of canonical and micro-canonical ensembles in long-range-interacting systems.
}
\label{fig:ensemble_equivalence}
\end{figure}

In the present paper, we consider an open question about the equivalence between canonical and micro-canonical ensembles (Fig.~\ref{fig:ensemble_equivalence}),
including systems with long-range interactions
(see also outlook in the review~\cite{kliesch2018properties}).
A microcanonical ensemble describes the state distribution of an isolated system with fixed total energy, while a canonical ensemble characterizes the state distribution of a system connected to a heat bath at a fixed temperature.
The equivalence of these two types of ensemble has been studied over a long time as a fundamental subject in statistical mechanics.
The traditional studies on the ensemble equivalence focus on the thermodynamic functions~\cite{ruelle1999statistical,doi:10.1063/1.1665684, lima1971equivalence,Lima1972,Georgii1995,de2006quantum,Touchette2015}.
More recently, the ensemble equivalence has been further extended to expectation values for arbitrary local observables~\cite{Muller2015,brandao2015equivalence,Tasaki2018}.
In such a generalization, there are many open problems especially on the finite-size effect for the error between the two ensembles.

The problem of ensemble equivalence can be classified roughly into three categories: 
i) conditions where the two ensembles are equivalent in the thermodynamic limit, 
ii) quantitative estimation of the difference between the two ensembles for a fixed system size,
iii) the possible widths of the energy shell in a micro-canonical ensemble as a function of the system size.
As for the problem i), extensive studies have been published both in classical~\cite{ruelle1999statistical,doi:10.1063/1.1665684, Georgii1995,Touchette2015} and quantum many-body systems~\cite{lima1971equivalence,Lima1972,de2006quantum,Muller2015}. 
More recently, regarding the problem ii), the finite-size effect on ensemble equivalence was considered explicitly in Refs.~\cite{brandao2015equivalence,Tasaki2018}.
For an arbitrary observable, the difference between the averages of the canonical and the micro-canonical ensembles has been quantitatively determined under the assumption of clustering (i.e., exponential decay of bipartite correlations).
So far, state of the art analyses of this problem estimate the difference as $\orderof{n^{-1/4}}$~\cite{Tasaki2018} with $n$ being system size under the assumptions of clustering and rapid convergence of the Massieu function.
Finally,  the problem iii) is raised as an open question that is relevant to the eigenstate thermalization hypothesis (ETH)~\cite{brandao2015equivalence}.
In other words, if one chooses an arbitrarily small energy width $\Delta$, even a single eigenstate (i.e., $\Delta\to +0$) is equivalent to the canonical ensemble.
However, the ETH is known to be violated in integrable systems, so the energy width is in fact limited unless some specific properties of the dynamics are assumed~\cite{doi:10.1080/00018732.2016.1198134,doi:10.1146/annurev-conmatphys-031214-014726}.

This paper aims to derive a non-trivial lower bound on the energy width of generic models without assuming specific dynamical properties of the system.
We note that it is already known that the energy width can be as small as $\orderof{n^{-1/2+\eta}}$ ($\eta >0$) for short-range-interacting spin systems~\cite{Tasaki2018}). The analysis given below goes beyond this estimation with a cluster-expansion analysis of the generic models.

In long-range-interacting systems, ensemble equivalence can
be violated~\cite{PhysRevLett.87.030601,CASETTI2007318}.
Considering this, we aim to identify the conditions under which ensemble equivalence is reliably ensured.
When analyzing the ensemble equivalence, we need to discuss the properties of the canonical state (i.e., the Gibbs state or the thermal-equilibrium state) at finite temperatures:
\begin{eqnarray}
\rho=\frac{e^{-\beta H}}{Z} \label{Gibbs_intro_1}
\end{eqnarray}
with $Z:=\tr (e^{-\beta H})$, where $H$ and $\beta$ are the system’s Hamiltonian and the inverse temperature, respectively.
At temperatures above a critical threshold (or in high-temperature phases), the clustering property has been proven in both classical systems~\cite{Gross1979} and quantum-spin systems~\cite{Araki1969,Park1995,ueltschi2004cluster, PhysRevX.4.031019,frohlich2015some} with short-range interactions.
However, long-range-interacting systems do not usually have a finite correlation length at any temperature, so we need to rely on a property other than the clustering.

In the present paper, we use the concentration bound as the basis of our analysis. 
If the spins are independent of each other, the following Chernoff--Hoeffding concentration inequality~\cite{ref:Chernoff,ref:Hoeffding_ineq} is known to hold.  
Roughly speaking, this inequality states that the probability distribution for a macroscopic observable is concentrated tightly around the average value. 
Let us consider a product state $\rho_0=\rho_{1}\otimes\rho_{2} \otimes \cdots \otimes \rho_{n}$ of an $n$-spin system.  
Then, the Chernoff--Hoeffding inequality upper-bounds the probability distribution of a one-body observable $A=\sum_{i}^{n} a_{i}$ with $\|a_i\|=1$ in Gaussian form:
\begin{eqnarray}
 \int_{x_0+\langle A \rangle}^\infty \tr [\rho_0 \delta(A-x)] dx \le \exp\left[-C \left( \frac{x_0}{\sqrt{n}} \right)^\gamma \right] \quad (x_0>0),\label{ineq:Chernoff_bound}
\end{eqnarray}
with $\gamma=2$, where $\delta(x)$ is the delta function, $\langle A \rangle:=\tr (\rho_{0} A)$ and $C$ is a constant that does not depend on the system size $n$.
We are concerned with whether the inequality~\eqref{ineq:Chernoff_bound} holds beyond the setup of product states and one-body observables.
In weakly correlated spin systems, inequality~\eqref{ineq:Chernoff_bound} has been generalized in several ways.
First, for product states or short-range entangled states (see \cite{ref:Wen-topo} for the definition), inequality~\eqref{ineq:Chernoff_bound} with $\gamma=2$ has been proven for probability distributions of generic few-body observables~\cite{Kuwahara_2016,Anshu_2016}.
If we consider more general classes of states, the concentration inequality can be derived less strictly (i.e., $\gamma<2$): 
$\gamma=1$ for gapped ground states~\cite{Kuwahara_2016_gs,Kuwahara_2017} and $\gamma=1/(D+1)$ ($D$: the system dimension) for states with clustering~\cite{Anshu_2016}.
In these works, the locality of interactions in 
the Hamiltonian plays a central role~\cite{arad2014connecting}. 
Moreover, if we restrict the analysis to classical spin systems with short-range interactions, the concentration inequalities have been extensively investigated~\cite{Kulske2003,Chazottes2007,redig2009,Chazottes2017} at both high temperatures ($\gamma=2$) and low temperatures ($\gamma<2$).

In this paper, through the cluster expansion,  we derive the Gaussian concentration bound, for a generic many-body systems including long-range systems. Below, we list our findings in this paper: 
\begin{enumerate}
\item{} The Gaussian concentration inequality $(\gamma=2)$ is rigorously proven for long-range-interacting systems above a threshold temperature (see Corollary~\ref{main_corol_CH_inequality}).
\item{} Under the assumption that the concentration bound applies, we quantitatively prove the equivalence of canonical and micro-canonical ensembles (see Theorem~\ref{main_theorem_Ensemble Equivalence}).
\item{} By applying Theorem~\ref{main_theorem_Ensemble Equivalence} to high-temperature Gibbs states, the difference between the canonical and micro-canonical ensembles is bounded from above by ${\displaystyle \left[n^{-1} \log (n^{3/2}\Delta^{-1})\right]^{1/2}}$. Ensemble equivalence holds for sufficiently large systems (or $n\gg1$) as long as $\Delta=\exp(-n^{1-\eta})$ with $\eta>0$. 
\end{enumerate}

The above three results solve the problems i) to iii) accurately in the viewpoint of the system-size dependence.
For problem i), ensemble equivalence in long-range-interacting systems is rigorously proven above a threshold temperature (see Eq.~\eqref{eq:critical_temperature} for the specific value). 
For problem ii), the quantitative difference in the averages of the canonical and the micro-canonical ensembles is bounded by $\orderof{n^{-1/2}}$ up to a logarithmic correction for $\Delta=1/{\rm poly}(n)$.
Finally, for problem iii), ensemble equivalence holds approximately even for the energy width of $\Delta=e^{-\orderof{n}}$ (see Corollary~\ref{corol:main_theorem_Ensemble Equivalence}).
Because the density of states in energy spectrum is $e^{\orderof{n}}$ at most,  
the energy gap smaller than $e^{-{\cal O}(n)}$ implies that the individual eigenstates become visible and affect the ensemble equivalence. 
We note that the realization of ensemble equivalence for the infinitesimal limit of energy width leads to the ETH. However, the ETH cannot be proven without imposing specific properties such as the non-integrability of the system~\cite{doi:10.1080/00018732.2016.1198134,doi:10.1146/annurev-conmatphys-031214-014726}. Hence, it is plausible that one cannot reduce the energy gap smaller than $e^{-{\cal O}(n)}$ in the present general framework.
We thus conclude that our estimation for the limitation to the energy width is qualitatively precise. 


  The rest of this paper is organized as follows. In section 2, we explain the setup of our analysis and our main findings about the concentration bound using the cluster expansion.
  In section 3, we apply our findings to ensemble equivalence and the weak version of the ETH. In section 4, we discuss future perspectives. In section 5, we outline the mathematical structure we used to derive the results.

\section{Setup and Main results}

We consider a quantum spin system with $n$ spins, where each of the spins has $d$-dimensional Hilbert space. 
We let $V=\{1,2,3,\ldots,n\}$ be the whole set of spins, and we denote the local Hilbert space by $\mathcal{H}^v$ ($v\in V$) with $\dim (\mathcal{H}^v)=d$. 
Now, the total Hilbert space is given by $\mathcal{H}:=\bigotimes_{v\in V} \mathcal{H}^v$ with $\mathcal{D}_{\mathcal{H}}:= \dim (\mathcal{H}) =d^n$.
We define the space of linear operators on $\mathcal{H}$ as $\ban(\mathcal{H})$.
To characterize the interactions between spins, we write the system Hamiltonian $H \in \ban(\mathcal{H})$ as 
\begin{align}
H= \sum_{|\ed| \le k} h_\ed  ,
\label{eq:ham_graph}
\end{align}
where each of $\{h_\ed\}_{|\ed| \le k}$ denotes an interaction between the spins in $\ed \subset V$.
The Hamiltonian~\eqref{eq:ham_graph} describes a generic $k$-body-interacting system.
We define $\mathcal{E}$ as the set of all eigenstates and describe each of the energy eigenstates as $\ket{ E }\in \mathcal{E}$ such that $H\ket{ E }= E \ket{ E }$.

We next consider the Hamiltonian for which the spectrum of the local Hamiltonian is finite. More precisely, we impose the condition 
\begin{align}
\sum_{\ed : \ed \ni v}\|h_{\ed} \| \le g  \for \forall v\in V ,\label{eq:extensiveness}
\end{align}
where $\|\cdots\|$ is the operator norm and $\sum_{\ed : \ed \ni v}$ sums up all the interactions that involve the spin $v$. 
We can directly obtain the following inequality for the total norm of the Hamiltonian:
\begin{align}
\|H\| \le \sum_{v\in V} \sum_{\ed : \ed \ni v}\|h_{\ed} \| \le \sum_{v\in V} g  = g |V| =gn \label{g-extensive_Ham}
\end{align}
Thus, inequality~\eqref{eq:extensiveness} upper-bounds the one-spin energy by $g$.

The above class of Hamiltonians includes long-range-interacting spin systems with a power-law decay on a lattice along with the short-range-interacting case.
For example, let us consider the following Hamiltonian of a $D$-dimensional lattice system that has interactions with a power-law decay of $1/r^\alpha$ ($r$: interaction length):
\begin{align}
H=\frac{1}{\tilde{N}}\sum_{i,j} \frac{J}{r_{i,j}^\alpha} (\sigma_i^x \sigma_j^x + \sigma_i^y \sigma_j^y + \sigma_i^z \sigma_j^z ),
\end{align}
with $J=\orderof{1}$, where $r_{i,j}$ is the Manhattan distance between spins $i$ and $j$ defined by the lattice geometry and $\tilde{N}$ is determined  so that the finite norm~\eqref{eq:extensiveness} of the local Hamiltonian is satisfied.
If the exponent $\alpha$ is greater than $D$, we have $\tilde{N}=\orderof{1}$.
On the other hand, for $\alpha\le D$, we need to consider that $\tilde{N}=\orderof{n^{D-\alpha}}$ due to condition \eqref{eq:extensiveness}.
In this example, we have $k=2$ in Eq.~\eqref{eq:ham_graph}.
This type of the interaction contains the Blume--Emery--Griffiths (BEG) model with infinite-range interactions (i.e., $\alpha=0$). 
In this model, the ensemble inequivalence has been previously investigated at low temperatures~\cite{PhysRevLett.87.030601}.
On the other hand, our result on the ensemble equivalence in Sec.~\ref{sec:Ensemble Equivalence} is applied to high-temperatures as in \eqref{eq:critical_temperature} and does not contradict the results in~\cite{PhysRevLett.87.030601}.
Moreover, it is noteworthy that the Hamiltonian~\eqref{eq:ham_graph} can also apply to quantum systems within infinite-dimensional networks, in which the breaking of ensemble equivalence has been reported~\cite{PhysRevLett.115.268701}.

Throughout this paper, we consider the Gibbs state of the Hamiltonian $H$ with inverse temperature $\beta$ as follows:
\begin{align}
\rho:= \frac{1}{Z}e^{-\beta H}, \quad Z:= \tr (e^{-\beta H}). \label{def:canonical_state}
\end{align}
Here, we aim to prove the following theorem below a certain threshold $\beta < \beta_c$, where $\beta_c$ does not depend on the system size $n$, but only on $k$ and $g$.  
\begin{theorem} \label{main_theorem_CH_inequality}
Let $\Fp \in \ban(\mathcal{H})$ be an arbitrary operator subject to the same conditions as \eqref{eq:extensiveness}, namely
\begin{align}
\Fp := \sum_{|\ed| \le k} f_\ed \quad {\rm with} \quad \sum_{\ed: \ed \ni v}\| f_\ed \|  \le g  \for v\in V.   \label{Def:Omega_op}
\end{align}
Then, if the inverse temperature satisfies 
\begin{align}
\beta < \beta_c := \frac{1}{8e^3 gk}, \label{eq:critical_temperature}
\end{align}
the Gibbs state $\rho$ satisfies the following inequality:
\begin{align}
  \log \left[\tr \left(e^{
  -\tau \Fp
  } \rho\right) \right] \le -\tau \langle \Fp \rangle_{\beta}  +\frac{\tau^2 \bar{\Fp}}{\beta_c-\beta-|\tau|}  ,   \label{eq:CH_inequality}
\end{align}
where 
\begin{align}
\langle \Fp \rangle_{\beta} := \tr(\Fp \rho)
\end{align}
 and we assume $|\tau| < \beta_c - \beta$ and $\bar{\Fp}$ are defined as  $\bar{\Fp} := \sum_{|\ed |\le k}\| f_\ed \|$.
\end{theorem}

For the sake of a clear presentation, we give the proof in the section~\ref{proof_of_Ch_ineq} and next we discuss several physical applications of the theorem.

This theorem implies the following Chernoff--Hoeffding inequality:
\begin{corol} \label{main_corol_CH_inequality}
We assume the conditions of Theorem~\ref{main_theorem_CH_inequality} and let $P_\rho(x)$ be
\begin{align}
P_\rho(x):= \tr[ \rho \delta (x-\Fp)]
\end{align}
with $\delta (x)$ being the delta function. We then obtain
\begin{align}
P_\rho(|x-\langle \Fp \rangle_{\beta}|\ge x_0) &:= \int_{|x-\langle \Fp \rangle_{\beta}|\ge x_0}  P_\rho(x) dx \le 2\exp\left( -\frac{x_0^2}{c_\beta \bar{\Fp}}  \right) \label{ineq:Chernoff_x_corol}
\end{align}
where we define
\begin{align}
c_\beta:=\frac{2}{\beta_c-\beta}.
\end{align}
\end{corol}

We next compare the above concentration inequality with findings from the literature. 
Around the average value $x_0=\orderof{n^{1/2}}$, the well-known central-limit theorem has been derived for several classes of quantum systems with translational invariance~\cite{ref:Hartmann,ref:vets,ref:vets2,matsui2002bosonic}.
This theorem states that the distribution of a macroscopic observable is not only bounded by Gaussian function, but it also converges to a Gaussian normal distribution in the thermodynamic limit ($n\to \infty$). The more-refined statement of the Berry-Essen theorem~\cite{ref:Berry,ref:Essen} has been proven for arbitrary quantum states that have the property of clustering~\cite{brandao2015equivalence}.  
Both the above theorems impose a stronger limitation than inequality~\eqref{ineq:Chernoff_x_corol} in that they prove exact convergence to the Gaussian distribution in the limit $n\to \infty$.
On the other hand, for finite $n$, these theorems cannot give the tight asymptotic behavior of the tail of probability distribution; indeed, the optimal convergence behavior is $\orderof{1/\sqrt{n}}$ at most, as found in Ref.~\cite{brandao2015equivalence}.

As for the asymptotic behavior of finite systems for $x_0=\orderof{n}$, various studies have addressed the large deviation~\cite{Touchette20091,Ogata2010,Large-diviation_hiai,ref:Marco_LD,Netocny2004}.
The large-deviation theorem asserts that the probability
becomes exponentially small as the system size $n$ increases:
\begin{align}
  &P_\rho(|x-\langle \Fp \rangle_{\beta}|\ge x_0) = \exp\left[ -n I (x_0/n)    + \orderof{n^{1-\kappa} } \right] \label{ineq:large-deviation}
\end{align}
with $\kappa>0$, where the rate function $I (\cdot)$ is non-zero and smooth.
The large-deviation theorem is stronger than the Chernoff--Hoeffding inequality~\eqref{ineq:Chernoff_x_corol} since it gives the correct asymptotic
exponential decay in the probability for $x_0=\orderof{n}$.
However, the large-deviation
theory is focused on the large-deviation function $I (x_0/n)$ and does not discuss
the rate of decay around the average value due to the sub-leading term of $\orderof{n^{1-\kappa}}$
that is written in Eq.~\eqref{ineq:large-deviation}.
This
aspect is crucial when discussing how finite size affects the equivalence between the canonical and micro-canonical distributions (see Sec.~\ref{sec:Ensemble Equivalence}).

\textit{Proof of Corollary~\ref{main_corol_CH_inequality}.}
Without loss of generality, we here set 
$\langle \Fp \rangle_{\beta}=\tr(\Fp \rho)=0$ and $\tau \le (\beta_c-\beta)/2$, and inequality~\eqref{eq:CH_inequality} then reads 
\begin{align}
\tr \left(e^{-\tau \Fp  } \rho\right) \le e^{c_\beta \tau^2 \bar{\Fp}} .
\end{align}
Using the above inequality, we obtain for $0< \tau < \beta_c-\beta$ and $x\ge 0$
\begin{align}
P_\rho(x\ge x_0) =\int_{x\ge x_0} \tr[ \rho \delta (x-\Fp)]dx
&= \int_{x\ge x_0} \tr[ \rho e^{\tau \Fp  } e^{-\tau \Fp  } \delta (x-\Fp)]dx           \notag \\
&\le   e^{-\tau x_0} \cdot \tr  \left(e^{\tau \Fp  }\rho\right) \notag \\
&\le e^{-\tau x_0+c_\beta \tau^2 \bar{\Fp} }
=\exp\left(c_\beta \bar{\Fp} \left[\tau - x_0/ (c_\beta \bar{\Fp})\right]^2 -x_0^2 /(c_\beta \bar{\Fp})  \right). \label{Ch_moment_prob_esti}
\end{align}
By using $\bar{\Fp} \ge \| \Fp\| \ge x_0$, we have 
\begin{align}
\frac{x_0}{c_\beta \bar{\Fp}} \le \frac{1}{c_\beta}=  \frac{\beta_c-\beta}{2}.
\end{align}
Thus, $\tau$ can be chosen as $\tau=x_0/ (c_\beta \bar{\Fp}) \le (\beta_c-\beta)/2$ in \eqref{Ch_moment_prob_esti} and we obtain inequality~\eqref{ineq:Chernoff_x_corol} for $x_0\ge 0$.
We can prove the case of $x_0\le 0$ in the same way. 
This completes the proof of Corollary~\ref{main_corol_CH_inequality}. $\square$

The Chernoff--Hoeffding inequality~\eqref{ineq:Chernoff_x_corol} also gives information about the density of
  states:
\begin{corol} \label{main_corol_Density_inequality}
For an arbitrary few-body Hamiltonian like that of Eq.~\eqref{eq:ham_graph} with \eqref{eq:extensiveness}, 
the total number of energy eigenstates in $ E  \in [-x_0, x_0 ]$ is bounded from below by
\begin{align}
\frac{\# \left\{ \ket{ E } \in \mathcal{E} |  E  \in [-x_0, x_0 ] \right\}}{d^n} \ge 1-2 \exp\left( -\frac{x_0^2}{c_0 g n}  \right) ,  \label{main_corol_Density_inequality_statement}
\end{align}
where $c_0=2/\beta_c$ and we set $\tr (H)=0$. 
\end{corol}
\textit{Remark.} This corollary does not characterize the Gibbs states but instead relates to the Hamiltonian itself. 
This analysis rigorously proves that the density of energy eigenstates follows the Gaussian concentration bound as the temperature goes to infinity.
Thus, the spectral distributions of \textit{all} few-body Hamiltonians resemble those of one-body Hamiltonians. 
This result is a generalization of the Keating’s proof ~\cite{Keating2015} (see Theorem~2 in the references) that the Gaussian concentration~\eqref{main_corol_Density_inequality_statement} holds for translation-invariant spin chains.

\textit{Proof of Corollary~\ref{main_corol_Density_inequality}.}
By choosing the infinite-limit temperature states (i.e., $\beta=0$) in Corollary~\ref{main_corol_CH_inequality}, we find $\rho=\hat{1}/d^n$ and $\Fp=H$
\begin{align}
\frac{1}{d^n} \int_{|x -\tr(H)/d^n | \ge x_0} \tr[\delta(H-x) ]dx \le 2 \exp\left( -\frac{x_0^2}{c_0 \bar{H}}  \right) .
\label{ineq:Chernoff_x_corol_2}
\end{align}
Note that $\langle H \rangle_{\beta} = \tr(H)/d^n$ for $\beta=0$. 
Applying the condition $\tr(H)=0$, we have
\begin{align}
\# \left\{ \ket{ E } \in \mathcal{E} |  E  \in [-x_0, x_0 ] \right\} = \int_{|x| \le x_0}\tr[\delta(H-x) ]dx  =  1- \int_{|x| > x_0}\tr[\delta(H-x)] dx     \label{ineq:Chernoff_x_corol_2_density}
\end{align}
and the condition \eqref{eq:extensiveness} gives 
\begin{align}
\bar{H} := \sum_{|\ed| \le k}\|h_\ed \| \le gn .\label{ineq:Chernoff_x_corol_2_hbar}
\end{align}
Then, by applying Eq.~\eqref{ineq:Chernoff_x_corol_2_density} and inequality~\eqref{ineq:Chernoff_x_corol_2_hbar} to \eqref{ineq:Chernoff_x_corol_2}, 
we obtain the inequality~\eqref{main_corol_Density_inequality_statement} under the condition $\tr(H)=0$. 
This completes the proof. $\square$

\section{Concentration bound and Equivalence of the canonical and the micro-canonical distributions} \label{sec:Ensemble Equivalence}

We here consider the equivalence of the canonical and micro-canonical distributions. 
Following the setup discussed in Refs.~\cite{brandao2015equivalence,Tasaki2018}, we first define the canonical and micro-canonical averages of an arbitrary operator $O\in \ban(\mathcal{H})$ as follows:
\begin{align}
&\langle O \rangle_{\beta} := \frac{1}{Z} \tr ( O e^{-\beta H}), \label{cb} \\
&\langle O\rangle_{U,\Delta} :=\frac{1}{\mathcal{N}_{U,\Delta}} \sum_{ E  \in ( U-\Delta, U ] } \bra{ E } O \ket{ E }, \label{def:cano_micro}
\end{align}
  where (\ref{cb}) and (\ref{def:cano_micro}) are the averages of observable $O$ over the canonical and micro-canonical ensembles, respectively. 
The quantity $\mathcal{N}_{U,\Delta}$
is the total number of energy eigenstates in $ E  \in (  U-\Delta, U ]$, namely,
\begin{align}
\mathcal{N}_{U,\Delta} := \tr \left( \sum_{ E  \in (  U-\Delta, U ] } \ket{ E } \bra{ E } \right). 
\end{align}
To characterize the micro-canonical ensemble, we choose $U$ as 
\begin{align}
U = \delta \nu^\ast  ,\quad  \nu^\ast:= {\rm argmax}_{\nu \in \mathbb{Z}}\left( e^{-\beta \nu \Delta }\mathcal{N}_{\nu \delta ,\delta}\right), \quad \delta := \min(\Delta, 1/\beta) \label{Def:U_micro_en}
\end{align}
If $\Delta \le 1/\beta$ the
width of energy shell $\Delta$ is equal to $\delta$.
In the standard formulation of the micro-canonical ensemble, the energy $U$ is fixed arbitrarily and the choice of \eqref{Def:U_micro_en} is not standard; 
for example, in Ref.~\cite{brandao2015equivalence}, $U$ was defined as $\tr(\rho H)$.
However, we adopt the choice~\eqref{Def:U_micro_en} following Ref.~\cite{Tasaki2018} in order to apply the proof techniques therein.

We are interested in the difference between the canonical average $\langle \Fp \rangle_{\beta}$ and the micro-canonical average $\langle \Fp \rangle_{U,\Delta}$.
For this purpose, we aim to prove that almost all the eigenstates in the energy shell $( U-\Delta, U]$ have the same expectation value as $\langle \Fp  \rangle_{\beta}$. 
We consider the probability distribution $P_{U,\Delta}(x)$ such that
\begin{align}
P_{U,\Delta}(x) = \frac{1}{\mathcal{N}_{U,\Delta}} \sum_{ E  \in ( U-\Delta, U ] }  \delta (x -  \bra{ E } \Fp  \ket{ E }) .\label{def:P_omega_x}
\end{align}
We now aim to derive the upper bound on the cumulative probability distribution as 
\begin{align}
P_{U,\Delta}(|x| \ge x_0) := \int_{x_0}^\infty P_{U,\Delta}(x)  dx + \int_{-\infty}^{x_0} P_{U,\Delta}(x)  dx.
\end{align}

Based on a concentration bound like \eqref{ineq:Chernoff_bound}, we can prove the following theorem:
\begin{theorem} \label{main_theorem_Ensemble Equivalence}
Let $\Fp\in \ban(\mathcal{H})$ be a few-body operator as in Eq.~\eqref{Def:Omega_op}.
Under the assumption that a Gibbs state \eqref{def:canonical_state} satisfies the following concentration bound for $\Fp$ such that
\begin{align}
  P_\rho(|x-\langle \Fp \rangle_\beta|\ge x_0)
  =
\int_{|x - \langle \Fp \rangle_{\beta}| \ge x_0} P_{\rho} (x) dx 
  \le \exp\left[- \left( \frac{x_0}{\sqrt{\tilde{c} g n}} \right)^\gamma \right] 
\label{ineq:Chernoff_x_assump}
\end{align}
with $\gamma$ and $\tilde{c}$ a positive constant of $\orderof{1}$,  we have 
\begin{align}
P_{U,\Delta}(|x-\langle \Fp \rangle_\beta| \ge x_0) \le  C_\delta \exp \left[ - \gamma \left( \frac{|x_0-\langle \Fp \rangle_{\beta}|}{\sqrt{\tilde{c}e gn }} \right)^{\gamma}  \right] , \label{CH_micro_canonical}
\end{align}
where
\begin{align}
C_\delta := \frac{16e^2 \sqrt{g n}}{\gamma\sqrt{\tilde{c}} }\left (1+\frac{gn}{\delta} \right) = \orderof{\delta^{-1}n^{3/2}}.
\end{align}
\end{theorem}
This theorem implies the following corollary:
\begin{corol} \label{corol:main_theorem_Ensemble Equivalence}
Under the assumption in Theorem~\ref{main_theorem_Ensemble Equivalence}, we have
\begin{align}
\frac{1}{n}|\langle \Fp \rangle_{U,\Delta}   - \langle \Fp \rangle_{\beta} | \le C_2\frac{ \log^{1/\gamma} \bigl(\delta ^{-1}n^{3/2}\bigr) }{\sqrt{n}} , 
\label{main_corol_Ensemble Equivalence}
\end{align}
with $C_2$ being a constant that depends only on the parameters $g$, $\gamma$ and $\tilde{c}$.
For arbitrary $\epsilon >\orderof{n^{-1/2}}$, we have
$\frac{1}{n}|\langle \Fp \rangle_{U,\Delta}   - \langle \Fp \rangle_{\beta} | \le \epsilon $
as long as  ${\displaystyle \delta \ge  \exp\left[-C ( \epsilon^2 n)^{2/\gamma} \right]}$ with $C=\orderof{1}$.
\end{corol}

Before giving the proof, we introduce the following useful lemma that was proven in a previous publication~\cite{PhysRevLett.124.200604}.
\begin{lemma} \label{Prob:expectation}
Let $p(x)$ be an arbitrary probability distribution whose cumulative distribution is bounded from above:
\begin{align}
  P(| x-a |\ge x_0):= \int_{|x-a| \ge x_0}p(x) dx \le \min(1, e^{-x_0^\gamma / \sigma + x_1 }) \, , ~~~~~\gamma >0, ~\sigma >0 ,~x_0>0\, .
\end{align}
Subsequently, for arbitrary $k \in \mathbb{N}$, we obtain 
\begin{align}
\int_{-\infty}^\infty |x-a|^k  p(x) dx \le  (2 \sigma x_1)^{k/\gamma} + \frac{k}{\gamma} (2\sigma)^{k/\gamma} \Gamma(k/\gamma)   \label{ineq:Prob:expectation}
\end{align}
with $\Gamma (\cdot)$ as the gamma function.
\end{lemma}

{~}

\textit{Proof of Corollary~\ref{corol:main_theorem_Ensemble Equivalence}.} 
If we apply this lemma to probability~\eqref{CH_micro_canonical} with the parameter set as
\begin{align}
\{a, \gamma, \sigma, x_1,k\} = \{\langle \Fp \rangle_{\beta},\gamma, (\tilde{c} e g n)^{\gamma/2},  \log (C_\delta),1\},
\end{align}
we obtain inequality~\eqref{main_corol_Ensemble Equivalence}. This completes the proof. $\square$ 

{~}

This theorem has several interesting implications:
 \begin{enumerate}
\item{} Theorem~\ref{main_theorem_Ensemble Equivalence} does not necessarily assume that the system is at a high temperature. 
\item{}
  The theorem is concerned with the specific choice of operator $\Fp$, which satisfies the concentration inequality~\eqref{ineq:Chernoff_x_assump}.
\item{} If the Chernoff--Hoeffding inequality holds (i.e., $\gamma=2$), ensemble equivalence holds approximately for exponentially small energy widths as  $\delta = \exp (-C\epsilon^2 n)$.
\item{} If we consider a high-temperature Gibbs state with  $\beta < \beta_c := 1/(8e^3 gk)$, from Corollary~\ref{main_corol_CH_inequality}, assumption~\eqref{ineq:Chernoff_x_assump} holds for arbitrary few-body operators with $\tilde{c}=c_\beta$ and $\gamma=2$.  
Thus, from Corollary~\ref{main_corol_CH_inequality} and \ref{corol:main_theorem_Ensemble Equivalence}, we can prove ensemble equivalence for arbitrary long-range-interacting systems above the temperature threshold $\beta_c$.
\end{enumerate}

Applying this corollary to the high-temperature regime, we conclude that ensemble equivalence holds approximately even for exponentially small energy width $\Delta$.
At first glance, this conclusion contradicts counterexamples to the ETH such as many-body localization~\cite{doi:10.1146/annurev-conmatphys-031214-014726}, under which no single eigenstate has the thermal property. 
In our theorem, it is true that the energy width can be as small as $e^{-C \epsilon^2 n}$. 
However, in this energy shell, there is still an exponentially large number of eigenstates as $\mathcal{N}_{U,\Delta} e^{-C \epsilon^2 n}$, where $\mathcal{N}_{U,\Delta}$ is typically as large as the total dimension of the Hilbert space $d^n$.
In order to reduce the micro-canonical ensemble to a single eigenstate, we have to choose sufficiently small $\Delta$ that $\mathcal{N}_{U,\Delta} e^{-C \epsilon^2 n}=\orderof{1}$, 
but such a choice of the energy width no longer gives a non-trivial bound on $\frac{1}{n}|\langle \Fp \rangle_{U,\Delta}   - \langle \Fp \rangle_{\beta} |$.
The seeming contradiction can thus be resolved.

On the other hand, in low-energy regions, the energy density $\mathcal{N}_{U,\Delta}$ can be much smaller than the total dimension of the Hilbert space $d^n$. 
If the concentration bound holds at low-temperatures, even a single eigenstate can resemble the canonical ensemble.
It indeed occurs under the assumption of the clustering property at sufficiently small temperatures~\cite{PhysRevLett.124.200604}.   

\subsection{Weak eigenstate thermalization}

Before giving the proof of Theorem~\ref{main_theorem_Ensemble Equivalence}, we will briefly discuss weak eigenstate thermalization~\cite{PhysRevLett.105.250401,PhysRevLett.119.100601}.
Under the eigenstate thermalization hypothesis, \textit{all the eigenstates} in an energy shell have the same properties as the canonical state, while
the weak eigenstate thermalization hypothesis argues that \textit{most} of the eigenstates in an energy shell have the same properties.
As discussed in Ref.~\cite{PhysRevLett.119.100601}, we consider the variance of $\bra{ E } \Fp\ket{ E }$ in the energy shell: \begin{align}
\frac{1}{{\cal N}_{U,\Delta}} \sum_{  E  \in ( U-\Delta, U ]} \left(\frac{ \bra{ E } \Fp \ket{ E }}{n} - \frac{\langle \Fp\rangle_{U,\Delta}}{n} \right)^2 
\end{align}
If this variance approaches $0$ in the limit $n\to 0$, almost all the eigenstates have the same expectation value as the micro-canonical average $\langle \Fp/n\rangle_{U,\Delta}$. Our concern is the effect from finite-sized variance with respect to the system size $n$.

Next, we will prove the following corollary:
\begin{corol} \label{corol:main_weak ETH}
Under the assumption in Theorem~\ref{main_theorem_Ensemble Equivalence}, we have
\begin{align}
\frac{1}{{\cal N}_{U,\Delta}} \sum_{  E  \in ( U-\Delta, U ]} \left(\frac{ \bra{ E } \Fp \ket{ E }}{n} - \frac{\langle \Fp\rangle_{U,\Delta}}{n} \right)^2 
\le C'_2\frac{ \log^{2/\gamma} \bigl(\delta ^{-1}n^{3/2}\bigr) }{n} ,  \label{ineq:main:weak_ETH}
\end{align}
with $C'_2$ being a constant that depends only on the parameters $g$, $\gamma$ and $\tilde{c}$.
\end{corol}
Therefore, provided that $\Delta = 1/{\rm poly}(n)$ (i.e., $\delta = 1/{\rm poly}(n)$ from Eq.~\eqref{Def:U_micro_en}), this estimation gives the upper bound on the variance of $\Fp/n$ by $\orderof{\log^{2/\gamma}(n)/n}$.
Up to the logarithmic correction, this estimation is qualitatively sharp since recent calculations by Alba~\cite{PhysRevB.91.155123} indeed showed 
that a (1/2)-spin isotropic Heisenberg chain expresses the variance on the order of $\orderof{1/n}$.

\textit{Proof of Corollary~\ref{corol:main_weak ETH}.}
We first set $\langle \Fp \rangle_{\beta}=0$.
Using definition~\eqref{CH_micro_canonical}, we first obtain
\begin{align}
&\frac{1}{{\cal N}_{U,\Delta}} \sum_{  E  \in ( U-\Delta, U ]} \left(\frac{ \bra{ E } \Fp \ket{ E }}{n} - \frac{\langle \Fp\rangle_{U,\Delta}}{n} \right)^2   \notag \\
=&\frac{1}{n^2}  \int_{-\infty}^\infty x^2 P_{U,\Delta}(x)  dx - \frac{1}{n^2}\left( \int_{-\infty}^\infty x P_{U,\Delta}(x)  dx \right)^2 
\le \frac{1}{n^2} \int_{-\infty}^\infty x^2 P_{U,\Delta}(x)  dx 
\end{align}
Under this assumption, we obtain inequality~\eqref{ineq:Chernoff_x_assump} for $P_{U,\Delta}(x)$ with $\langle \Fp \rangle_{\beta}=0$.
Hence, we can utilize Lemma~\ref{Prob:expectation} by choosing the parameters 
\begin{align}
\{a, \gamma, \sigma, x_1,k\} = \{0,\gamma, (\tilde{c} e g n)^{\gamma/2},  \log (C_\delta),2\}.
\end{align}
Then, inequality~\eqref{ineq:Prob:expectation} gives \eqref{ineq:main:weak_ETH}.
This completes the proof. $\square$

\subsection{Proof of Theorem~\ref{main_theorem_Ensemble Equivalence}}
Throughout the following proof, we set $\langle \Fp \rangle_{\beta}=0$.
We begin with the $m$th moment function ($m$ is even):
\begin{align}
M_{U,\Delta}(m) := \int_{-\infty}^\infty x^m P_{U,\Delta}(x)  dx .\label{moment_dist_omega}
\end{align}
From the definition~\eqref{def:P_omega_x}, we have
\begin{align}
M_{U,\Delta}(m) &= \frac{1}{\mathcal{N}_{U,\Delta}} \sum_{ E  \in ( U-\Delta, U ] }  ( \bra{ E } \Fp \ket{ E } )^m  \notag \\
&\le \frac{1}{\mathcal{N}_{U,\Delta}} \sum_{ E  \in ( U-\Delta, U ] }    \bra{ E }\Fp^m \ket{ E }= \left\langle \Fp^m\right\rangle_{U,\Delta} .
\label{moment_dist_omega_MC}
\end{align}
For an arbitrary quantum state $\ket{\psi}$, we have 
\begin{align}
 ( \bra{\psi} \Fp \ket{\psi} )^m  \le \bra{\psi} \Fp^m \ket{\psi}
\end{align}
due to the convexity of $x^m$ ($m$ is even).

Second, we consider the following relation which will be proven in the subsequent subsection 
\begin{align}
\frac{\langle \tilde{O} \rangle_{U,\Delta}}{\langle \tilde{O} \rangle_{\beta}} \le 2e \left (1+\frac{gn}{\delta} \right). \label{mc_Can_difference_Z}
\end{align}
for arbitrary non-negative operators $\tilde{O}\in \mathcal{B}(\mathcal{H})$.
By choosing $\tilde{O}= \Fp^m \ge 0$, we have
\begin{align}
M_{U,\Delta}(m)   \le \langle \Fp^m \rangle_{U,\Delta}   
&\le  2e \left (1+\frac{gn}{\delta} \right)  \langle\Fp^m \rangle_{\beta} \notag \\
&\le    \frac{8e}{\gamma}\left (1+\frac{gn}{\delta} \right) (\tilde{c}gn)^{m/2}\left(\frac{m+1}{\gamma}\right)^{(m+1)/\gamma} ,  \label{mc_Can_difference_0}
\end{align}
where we have exploited the fact that the assumption \eqref{ineq:Chernoff_x_assump} implies 
\begin{align}
 \langle\Fp^m \rangle_{\beta}\le \frac{4(\tilde{c}gn)^{m/2}}{\gamma} \Gamma\left(\frac{m+1}{\gamma}\right) \le \frac{4(\tilde{c}gn)^{m/2}}{\gamma} \left(\frac{m+1}{\gamma}\right)^{(m+1)/\gamma} , 
\end{align}
with $\Gamma(\cdot)$ being the gamma function.

Then, by using inequality~\eqref{mc_Can_difference_0}, we obtain
\begin{align}
P_{U,\Delta}(x \ge x_0) = \int_{x_0}^\infty P_{U,\Delta}(x)  dx &\le \frac{1}{x_0^m} \int_{x_0}^\infty  x^m P_{U,\Delta}(x)  dx   \notag \\
&\le   \frac{8e}{\gamma}\left (1+\frac{gn}{\delta} \right) \left(\frac{\tilde{c}gn}{x_0^2}\right)^{m/2}\left(\frac{m+1}{\gamma}\right)^{(m+1)/\gamma}  \notag \\
&=\frac{8e}{\gamma}\left (1+\frac{gn}{\delta} \right) \left(\frac{x_0^2}{\tilde{c}gn}\right)^{1/2}\left[ \left(\frac{m+1}{\gamma} \right)^{2/\gamma}\frac{\tilde{c}gn}{x_0^2}\right]^{\frac{m+1}{2}}.  \label{up_bound_P_omega_x_x0}
\end{align}
We now choose 
\begin{align}
m+1 = \left \lfloor \gamma \left( \frac{x_0^2}{\tilde{c}e gn } \right)^{\gamma/2} \right \rfloor \quad {\rm or} \quad 1,
\end{align}
and inequality~\eqref{up_bound_P_omega_x_x0} reduces to
\begin{align}
P_{U,\Delta}(x \ge x_0) \le \frac{8e^2 \sqrt{g n}}{\gamma\sqrt{\tilde{c}} }\left (1+\frac{gn}{\delta} \right)
\exp \left[ - \gamma \left( \frac{x_0}{\sqrt{\tilde{c}e gn }} \right)^{\gamma}  \right],  \label{up_bound_P_omega_x_x02}
\end{align}
where we use $x_0^2/\tilde{c}gn \le gn/\tilde{c}$ since $x_0\le gn$. 
We thus obtain inequality~\eqref{CH_micro_canonical} by combining the cases of $P_{U,\Delta}(x \le - x_0) $. 
$\square$

\subsubsection{Proof of inequality~\eqref{mc_Can_difference_Z}} 

For this proof, we consider
\begin{align}
\langle \tilde{O} \rangle_{U,\Delta} &=\frac{1}{\mathcal{N}_{U,\Delta}} \sum_{ E  \in ( U-\Delta , U] } \bra{ E } \tilde{O} \ket{ E } \notag \\
&\le \frac{e^{\beta U}}{\mathcal{N}_{U,\Delta}} \sum_{ E  \in ( U-\Delta , U ] }e^{-\beta  E } \bra{ E } \tilde{O} \ket{ E }\notag \\
&\le \frac{Z e^{\beta U}}{\mathcal{N}_{U,\delta}} \sum_{ E  \in ( -\infty , \infty ) }\frac{1}{Z}  e^{-\beta  E } \bra{ E } \tilde{O} \ket{ E } \le \frac{Z e^{\beta U}}{\mathcal{N}_{U,\delta}}\langle \tilde{O} \rangle_{\beta},
\label{mc_Can_difference_Z2}
\end{align}
where $\mathcal{N}_{U,\Delta} \ge  \mathcal{N}_{U,\delta}$ because $\delta = \min(\Delta, 1/\beta) \le \Delta$. 

To bound $\frac{Z e^{\beta U}}{\mathcal{N}_{U,\delta}}$ from above, we consider
\begin{align}
Z =\sum_{ E  \in ( -\|H\| , \|H\| ] } e^{-\beta  E }
&\le   \sum_{\nu\in \mathbb{Z}: \nu \delta \in ( -\|H\|-\nu  ,\|H\| +\nu ]}\mathcal{N}_{\nu \delta, \delta} e^{-\beta \delta (\nu-1) }    \notag \\
&\le  e^{\beta \delta} \left (2+\frac{2 \|H\|}{\delta} \right) \max_{\nu\in \mathbb{Z}}\left( \mathcal{N}_{\nu \delta ,\delta} e^{-\beta \nu \delta }\right)  \notag \\
&\le   2e \left (1+\frac{gn}{\delta} \right) \mathcal{N}_{\nu^\ast \delta ,\delta} e^{-\beta \nu^\ast \delta } = 
2e \left (1+\frac{gn}{\delta} \right) \mathcal{N}_{U ,\delta} e^{-\beta U}, \label{upper_bound_tilde_Z}
\end{align}
where we have used inequality~\eqref{g-extensive_Ham}, $\delta = \min(\Delta, 1/\beta) $ and the definition of $U$ in Eq.~\eqref{Def:U_micro_en}. 
By combining inequalities~\eqref{mc_Can_difference_Z2} and \eqref{upper_bound_tilde_Z}, we obtain inequality~\eqref{mc_Can_difference_Z}. $\square$

\section{Summary and future perspectives}

This paper has rigorously analyzed the concentration bound and the equivalence of canonical and micro-canonical distributions in long-range-interacting systems.
The first theorem~\ref{main_theorem_CH_inequality} (or Corollary~\ref{main_corol_CH_inequality}) ensures that the Gaussian concentration inequality holds above a temperature threshold $\beta_c=1/(8e^3 gk)$ with $g$ and $k$ being the parameters of the Hamiltonian. 
We then connected the concentration bound to ensemble equivalence with Theorem~\ref{main_theorem_Ensemble Equivalence}. 
This theorem is not restricted to high-temperature Gibbs states so it can be applied to more-general cases.
When applying the theorem to the high-temperature Gibbs states, the Gaussian concentration bound implies  
$\frac{1}{n}|\langle \Fp \rangle_{U,\Delta}   - \langle \Fp \rangle_{\beta} | \le C_2 \left(n^{-1}\log \bigl(\Delta ^{-1}n^{3/2}\bigr)\right)^{1/2}$ 
for arbitrary few-body operators $\Fp \in \ban(\mathcal{H})$ with $\Delta$ ($\le 1/\beta$) being the width of the energy shell.
As shown in Corollary~\ref{corol:main_weak ETH}, we also proven the existence of weak eigenstate thermalization, namely that almost all the eigenstates in the energy shell have a value similar to the micro-canonical average.
The above results are a first theoretical step in the quantitative treatment of ensemble equivalence and weak eigenstate thermalization in generic quantum systems that undergo 
long-range interactions.

Several open questions remain.
First, the Gaussian concentration bound in Corollary~\ref{main_corol_CH_inequality} is applied to only systems with few-body observables as Eq.~\eqref{Def:Omega_op}. 
This class of observables accounts for almost all thermodynamic properties that could be of interest. 
However, to discuss the distance of traces between the reduced density matrices of canonical and micro-canonical states, we need to consider the summation of non-local operators
 (see Refs.~\cite{brandao2015equivalence,Tasaki2018,PhysRevLett.124.200604}): 
 \begin{align}
\tilde{\Fp}= \sum_{i=1}^{\tilde{n}}\tilde{ \Fp}_i  ,\quad \|\tilde{\Fp}_i\| \le 1,
\end{align}
where $\{\tilde{\Fp}_i\}_{i=1}^{\tilde{n}}$ is supported for large subsystems $B_i \subset V$ with $|B_i|\gg 1$ that are not overlapped with each other (i.e., $B_i \cap B_j=\emptyset$).
In this case, we still expect the Gaussian concentration inequality to be in the form of $\exp [-x^2/(c\tilde{n})]$ above a threshold temperature. 
Unfortunately, the present proof cannot be extended to this case.

The second open question is whether we can prove ensemble equivalence for long-range-interacting systems at low temperatures. 
We have already shown that the concentration bound~\eqref{ineq:Chernoff_x_assump} on Gibbs states is a sufficient condition for ensemble equivalence to hold.
We note that the bound should not hold universally since violation of ensemble equivalence has been reported in Refs.~\cite{PhysRevLett.87.030601,CASETTI2007318}.
Therefore, a central task of future work is to identify the conditions under which the concentration bound~\eqref{ineq:Chernoff_x_assump} holds at low temperatures.

Finally, further applications of the concentration bounds to other problems in statistical mechanics remain as an important future challenge.

\section{Proof of Theorem~\ref{main_theorem_CH_inequality}} \label{proof_of_Ch_ineq}

In the following, we give the proof of Theorem~\ref{main_theorem_CH_inequality}.
The proof consists of several propositions (Propositions~\ref{prop: connceted_cluster_CH_ineq}, \ref{prop:explicit_form_derivation_CH}, \ref{prop:bound_on_Der} and \ref{prop: convergence of cluster expansion}), whose proofs are shown in Appendices.

\subsection{Cluster notation}

We first define several basic terminologies.
We define $E_k$ as the set of  $X\subset V$ such that $|X|\le k$, namely
\begin{align}
E_k := \bigl \{X\subset V \bigl | |X| \le k \bigr\}.
\end{align}
We call a multiset $w=\{\ed_1,\ed_2,\ldots,\ed_{|w|}\}$ with $X_j \in E_k$ for $j=1,2,\ldots , |w|$ as ``cluster'', where $|w|$ is the cardinality of $w$ (i.e., the number of subsets in $w$).
Here, $w$ is an unordered list in which the same element can appear more than once.
Also, the notation $\oplus$ indicates the union of two multisets, for example $\{X_1,X_2,X_2\} \oplus \{X_1,X_1,X_3\}=\{X_1,X_1,X_1,X_2,X_2,X_3\}$.   
We denote by $\Cl_{m}$ the set of $w$ with $|w|=m$.
We define $V_w \subseteq V$ as 
\begin{align}
V_w := \ed_1 \cup \ed_2 \cup \cdots \cup \ed_{|w|}.
\end{align}

Also, we define the connected cluster as follows:  
 \begin{definition}{\rm  \bf (Connected cluster)}\label{Def:Connected_cluster}
For a cluster $w\in \Cl_{|w|}$, we say that $w$ is a connected cluster if 
there are no decompositions of $w=w_1 \oplus w_2$ ($w_1\neq \emptyset$, $w_2\neq \emptyset$) such that $V_{w_1} \cap V_{w_2}= \emptyset$.
We denote by $\Gc_m$ the set of the connected clusters with $|w|=m$.
\end{definition}

\begin{figure}
\centering
\subfigure[Case of $w\in \Gc_4^L$
]
{\includegraphics[clip, scale=0.42]{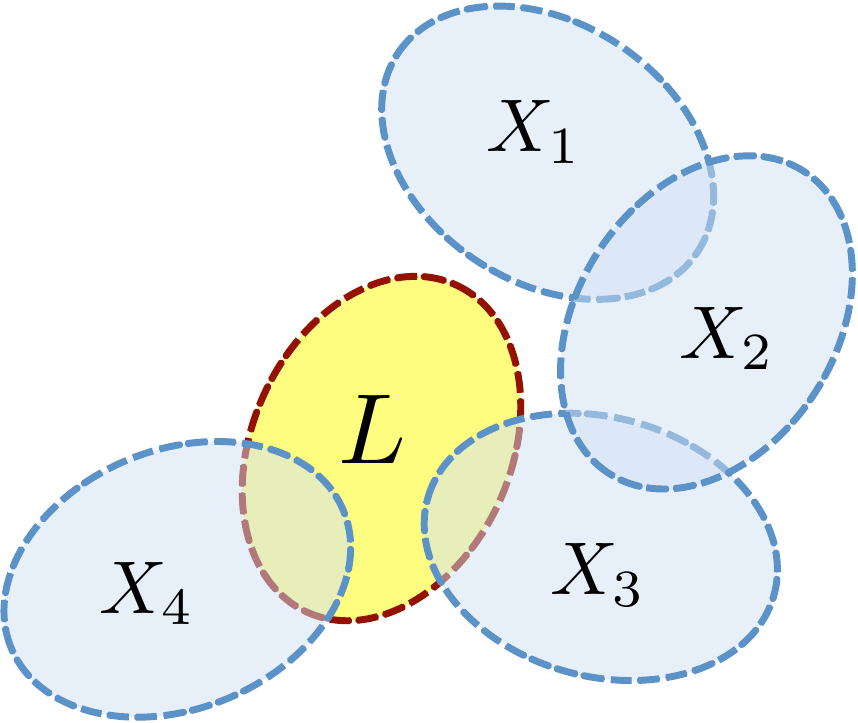}
}
\subfigure[Case of $w\notin \Gc_4^L$
]
{\includegraphics[clip, scale=0.42]{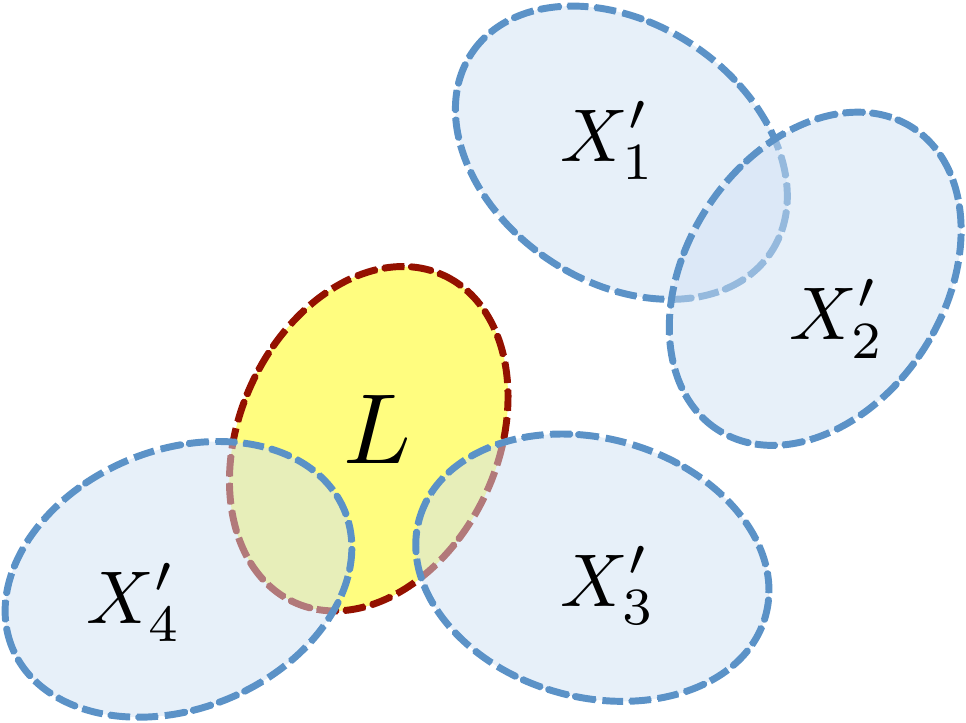}
}
\caption{Schematic pictures of clusters with $w \in \Gc_4^L$ and $w \notin \Gc_4^L$. Each of the elements $\{\ed_s | \ed_s \in E_k\}$ is a subset of the total set $V$ (i.e., $\ed \subset V$). In (a), there there are no decompositions of $w=w_1 \oplus w_2$ such that $(L\cup V_{w_1}) \cap V_{w_2}= \emptyset$ for $w=\{\ed_1,\ed_2,\ed_3,\ed_4\}$, whereas  in (b)
the decomposition $w'=w'_1 \oplus w'_2$ with $w'_1=\{\ed'_2,\ed'_3\}$ and $w'_2=\{\ed'_1,\ed'_4\}$ satisfies $(L\cup V_{w_1}) \cap V_{w_2}= \emptyset$.
}
\label{fig:connected_cluster}
\end{figure}

 \begin{definition}{\rm  \bf ($L$-connected cluster, FIG.~\ref{fig:connected_cluster})}\label{Def:Connected_cluster_to_L}
For any $L \subset V$ , we say that a cluster $w$ is $L$-connected when $\omega \oplus \{L\}$ is connected in the sense of Def.~\ref{Def:Connected_cluster};
that is, there are no decompositions of $w=w_1 \oplus w_2$ ($w_2\neq \emptyset$) such that $(L\cup V_{w_1}) \cap V_{w_2}= \emptyset$.
We denote by $\Gc_m^L$ the set of the connected clusters to $L$ with $|w|=m$. 
\end{definition}


\subsection{Cluster expansion} \label{cluster_expansion_CH_inequality}
We here introduce the cluster expansion to derive the moment generating function $\tr(e^{-\tau \Fp} \rho )$.
In order to treat the cluster expansion in simpler ways, we are going to use the following parametrization which has been also utilized in Ref.~\cite{PhysRevLett.124.220601}.
We first parametrize $H$ by a parameter set $\vec{a}:=\{a_\ed\}_{\ed\in E_k}$ as 
\begin{align}
H_{\vec{a}} = \sum_{\ed\in E_k} a_\ed h_\ed, \label{eq_parameterize_H}
\end{align}
where $H_{\vec{1}}= H$ with $\vec{1}=\{1,1,\ldots,1\}$.
By using Eq.~\eqref{eq_parameterize_H}, we define a parametrized Gibbs state $\rho_{\vec{a}}$ as  
\begin{align}
\rho_{\vec{a}} := \frac{e^{-\beta H_{\vec{a}}}}{Z_{\vec{a}}} 
\end{align}
where $Z_{\vec{a}}:= \tr(e^{-\beta H_{\vec{a}}})$.
Similarly, we parametrize $\Fp$ by $\Fp_{\vec{b}}$ as in Eq.~\eqref{eq_parameterize_H}; that is, $\Fp_{\vec{b}}=\sum_{\ed\in E_k} b_{\ed}f_{\ed}$.
In the cluster expansion which has been utilized in Ref.~\cite{PhysRevX.4.031019,frohlich2015some,Netocny2004}, they consider the Taylor expansion of $e^{-\beta H_{\vec{a}}}$ with respect to the parameters $\vec{a}$. 
Instead, we here utilize the Taylor expansion of $\log \left[\tr \left(e^{-\tau \Fp_{\vec{b}}} \rho_{\vec{a}}\right) \right]$ with respect to the parameters $\vec{a}$ and $\vec{b}$.

First, the Taylor expansion of $\rho_{\vec{1}}$ with respect to $\vec{a}$ reads
\begin{align}
\rho_{\vec{1}} = \sum_{m=0}^{\infty}\frac{1}{m!} \left[  \left( \sum_{\ed\in E_k}  \frac{\partial}{\partial a_\ed}  \right)^m \rho_{\vec{a}} \right]_{\vec{a}=\vec{0}} 
=\sum_{m=0}^{\infty}\frac{1}{m!} \sum_{\ed_1,\ed_2,\ldots,\ed_m\in E_k} \prod_{j=1}^m \frac{\partial}{\partial a_{\ed_j}} \rho_{\vec{a}} \Bigl|_{\vec{a}=\vec{0}},
\end{align}
where $\vec{0}=\{0,0,\ldots,0\}$.
By using the cluster notation, we obtain 
\begin{align}
\sum_{\ed_1,\ed_2,\ldots,\ed_m\in E_k} = \sum_{w \in \Cl_m} n_w ,
\end{align}
where $w=\{\ed_1, \ed_2 \ldots, \ed_m\}$ and $n_w$ is the multiplicity that $w$ appears in the summation.
For example, we have $n_w=12$ for $w=\{X_1,X_1,X_2,X_3\}$.
With this expression, one can write $\rho_{\vec{1}}$ as 
\begin{align}
\rho_{\vec{1}}  =& \sum_{m=0}^{\infty}\frac{1}{m!} \sum_{w \in \Cl_m} n_w \Der_{w}  \rho_{\vec{a}} \Bigl|_{\vec{a}=\vec{0}} \quad {\rm with} \quad \Der_{w}:= \prod_{j=1}^m \frac{\partial}{\partial a_{\ed_j}} . \label{Cluster_decomp_first}
\end{align}

We second expand the generating function $\log \left[\tr \left(e^{-\tau \Fp_{\vec{b}}} \rho_{\vec{1}}\right) \right]$ with respect to $\vec{b}$.
In the same way as the derivation of Eq.~\eqref{Cluster_decomp_first}, we obtain
\begin{align}
\log \left[\tr \left(e^{-\tau \Fp_{\vec{1}}} \rho_{\vec{1}}\right) \right]    
 =& \sum_{m=0}^{\infty}\frac{1}{m!} \sum_{\ed_1,\ed_2,\ldots,\ed_m\in E_k} \prod_{j=1}^m \frac{\partial}{\partial b_{\ed_j}}  \log \left[\tr \left(e^{-\tau \Fp_{\vec{b}} } \rho_{\vec{1}}\right) \right]\biggl| _{\vec{b}=\vec{0}} \notag \\
 =& \sum_{m=1}^{\infty}\frac{1}{m!} \sum_{\ed \in E_k}\frac{\partial}{\partial b_{\ed}}   \sum_{w\in \Cl_{m-1}} n_w \tilde{\Der}_{w}   \log \left[\tr \left(e^{-\tau \Fp_{\vec{b}} } \rho_{\vec{a}}\right) \right]\biggl| _{\vec{b}=\vec{0}} , \label{Cluster_decomp_second}
\end{align}
where we dropped the term with $m=0$ because of $\log[\tr(\rho_{\vec{1}})]=0$, and we defined
\begin{align}
\tilde{\Der}_{w} := \prod_{\ed \in w} \frac{\partial}{\partial b_{\ed}}.
\end{align}
The term with $m=1$ gives
\begin{align}
\sum_{\ed \in E_k}\frac{\partial}{\partial b_{\ed}}  \log \left[\tr \left(e^{-\tau \Fp_{\vec{b}} } \rho_{\vec{1}}\right) \right]\biggl| _{\vec{b}=\vec{0}}  =-\tau \sum_{\ed\in E_k}\tr (f_{\ed} \rho_{\vec{1}})= -\tau \tr (\Fp\rho_{\vec{1}})  ,
\end{align}
which reduces Eq.~\eqref{Cluster_decomp_second} to 
\begin{align}
&\log \left[\tr \left(e^{-\tau \Fp_{\vec{1}}} \rho_{\vec{1}}\right) \right]   +\tau \tr (\Fp\rho_{\vec{1}})   \notag \\
=&\sum_{m=2}^{\infty}\frac{1}{m!} \sum_{\ed \in E_k}\frac{\partial}{\partial b_{\ed}}   \sum_{w\in \Cl_{m-1}} n_w \tilde{\Der}_{w}   \log \left[\tr \left(e^{-\tau \Fp_{\vec{b}} } \rho_{\vec{1}}\right) \right]\biggl| _{\vec{b}=\vec{0}}  \notag \\
=&
\sum_{m_1=1}^{\infty}\sum_{\ed \in E_k}   \sum_{w_1\in \Cl_{m_1}}\frac{n_{w_1} \tilde{\Der}_{w_1\oplus \{X\}}}{(m_1+1)!}  
\sum_{m_2=0}^{\infty} \sum_{w_2 \in \Cl_{m_2}}\frac{n_{w_2}\Der_{w_2} }{m_2!}
   \log \left[\tr \left(e^{-\tau \Fp_{\vec{b}} } \rho_{\vec{1}}\right) \right]\biggl| _{\vec{a}=\vec{0},\vec{b}=\vec{0}}
,\label{Cluster_decomp_CH_ineq}
\end{align}
where we use the expansion~\eqref{Cluster_decomp_first} for $\rho_{\vec{1}}$ and $\tilde{\Der}_{w_1\oplus \{X\}}= (\partial/\partial b_{\ed})\tilde{\Der}_{w_1}$ in the second equation.

The expansion~\eqref{Cluster_decomp_CH_ineq} is only the multi-parameter Taylor expansion in itself.
However, we can prove that the summation with respect to $\sum_{w \in \Cl_m}$ reduces to quite a simple form by using the following proposition (see \ref{proof_prop: connceted_cluster_CH_ineq} for the proof):

\begin{prop} \label{prop: connceted_cluster_CH_ineq}
For arbitrary $w_1\in\Cl_{m_1}$ and $w_2\in\Cl_{m_2}$, we have 
 \begin{align}
\Der_{w_2} \tilde{\Der}_{w_1} \log \left[\tr \left(e^{-\tau \Fp_{\vec{b}}} \rho_{\vec{a}}\right) \right]\Bigl|_{\vec{a}=\vec{0},\vec{b}=\vec{0}} 
=0  \for w_1 \oplus w_2 \notin \Gc_{m_1+m_2} \label{DMCE_Cluster_decomp_CH_ineq0}.
\end{align}
\end{prop}
By applying the above proposition to the cluster expansion~\eqref{Cluster_decomp_CH_ineq}, we obtain
\begin{align}
&\log \left[\tr \left(e^{-\tau \Fp_{\vec{1}}} \rho_{\vec{1}}\right) \right]   +\tau \tr (\Fp\rho_{\vec{1}})   \notag \\
=&
\sum_{\ed \in E_k} \frac{\partial}{\partial b_{\ed}} 
\sum_{m_1=1}^{\infty} \sum_{m_2=0}^{\infty}  \sum_{\substack{w_1\in \Cl_{m_1}, w_2 \in \Cl_{m_2}\\ w_1\oplus w_2 \in \Gc_{m_1+m_2}^X}}\frac{n_{w_1}n_{w_2} }{(m_1+1)!m_2!}  
\tilde{\Der}_{w_1}  \Der_{w_2}   \log \left[\tr \left(e^{-\tau \Fp_{\vec{b}} } \rho_{\vec{1}}\right) \right]\biggl| _{\vec{a}=\vec{0},\vec{b}=\vec{0}}
,\label{DMCE_Cluster_decomp_CH_ineq}
\end{align}
where we use  the definition~\ref{Def:Connected_cluster_to_L}.

%
%

\subsection{Summation of the expansion} \label{cluster_expansion_cond_mutual_information}

In order to upperbound the summation with respect to connected clusters, the estimation of the upper bound of 
\begin{align}
\Der_{w_2} \tilde{\Der}_{w_1}  \frac{\partial}{\partial b_\ed}\log \left[\tr \left(e^{-\tau \Fp_{\vec{b}}} \rho_{\vec{a}}\right) \right]\Bigl|_{\vec{a}=\vec{0},\vec{b}=\vec{0}} .
\end{align}
 is crucial. 
 Because of $\frac{\partial}{\partial b_\ed}\log Z_{\vec{a}}=0 $,  we first obtain
\begin{align}
\frac{\partial}{\partial b_\ed}\log \left[\tr \left(e^{-\tau \Fp_{\vec{b}}} \rho_{\vec{a}}\right) \right] 
=-\tau  \frac{\tr \left( f_\ed e^{-\tau \Fp_{\vec{b}}} e^{-\beta H_{\vec{a}}} \right)} {\tr (e^{-\tau \Fp_{\vec{b}}} e^{-\beta H_{\vec{a}}})} =-\tau \tr \left( f_\ed \Phi_{\vec{a},\vec{b}}   \right), \label{der_b_e:phi_a_b} 
\end{align}
where $\Phi_{\vec{a},\vec{b}}$ is defined as
\begin{align}
\Phi_{\vec{a},\vec{b}} := \frac{e^{-\tau \Fp_{\vec{b}}} e^{-\beta H_{\vec{a}}}} {\tr (e^{-\tau \Fp_{\vec{b}}} e^{-\beta H_{\vec{a}}})}. \label{def:phi_a_b}
\end{align}

In the following proposition, we give an explicit form of $\Der_{w_2} \tilde{\Der}_{w_1} \tr \bigl( f_\ed \Phi_{\vec{a},\vec{b}}   \bigr)\bigl |_{\vec{a}=\vec{0},\vec{b}=\vec{0}}$ (see \ref{proof_prop:explicit_form_derivation_CH} for the proof).
\begin{prop} \label{prop:explicit_form_derivation_CH}
Let us take $m$ copies of the total Hilbert space $\mathcal{H}$ and consider $\mathcal{H}^{\otimes m+1}=\mathcal{H}_{1}\otimes  \mathcal{H}_{2}  \otimes \cdots \otimes\mathcal{H}_{m+1}$. 
We here introduce the following notation:
\begin{align}
\copyave{q}{\cdots} :=   \frac{\tr_{\mathcal{H}^{\otimes q}}(\cdots)}{\tr_{\mathcal{H}^{\otimes q}}(\hat{1})} \label{def:copyave}
\end{align}
for an arbitrary $q\in \mathbb{N}$.
Then, for an arbitrary cluster $w=w_1 \oplus w_2$, we have 
\begin{align}
&\Der_{w_2} \tilde{\Der}_{w_1} \tr \left( f_\ed \Phi_{\vec{a},\vec{b}}   \right) |_{\vec{a}=\vec{0},\vec{b}=\vec{0}}  \notag \\
=& \frac{(-\tau)^{m_1} }{m_1!} \frac{(-\beta)^{m_2}}{m_2!} 
\sum_{\sigma_1,\sigma_2} 
\copyave{m+1}{
f_\ed^{(0)} f_{\ed_{1,\sigma_1(1)}}^{(1)} f_{\ed_{1,\sigma_1(2)}}^{(2)} \cdots f_{\ed_{1,\sigma_1(m_1)}}^{(m_1)} 
h_{\ed_{2,\sigma_2(1)}}^{(m_1+1)} h_{\ed_{2,\sigma_2(2)}}^{(m_1+2)} \cdots h_{\ed_{2,\sigma_2(m_2)}}^{(m)} 
},
  \label{simple_expression_der_CH_ineq}
\end{align}
with $w_1=\{\ed_{1,j}\}_{j=1}^{m_1}$, $w_2=\{\ed_{2,j}\}_{j=1}^{m_2}$ and $m_1+m_2=m$, 
where for an arbitrary operator $O\in \ban(\mathcal{H})$ we define
\begin{align}
&O_{\mathcal{H}_s} :=\overbrace{\hat{1}\otimes \hat{1} \otimes  \cdots \otimes \hat{1}}^{s-1} 
\otimes  ~O                \otimes
\overbrace{\hat{1}\otimes \hat{1} \otimes  \cdots \otimes \hat{1}}^{m+1-s}\notag \\
& O^{(0)} :=O_{\mathcal{H}_1} ,\quad O^{(s)} := O_{\mathcal{H}_1} + O_{\mathcal{H}_2} + \cdots +O_{\mathcal{H}_s} -  s O_{\mathcal{H}_{s+1}}  , \label{extended_operator_prop}
\end{align}
for $ s=1,2,\ldots,m$. 
Also, $\sum_{\sigma_1}$ and $\sum_{\sigma_2}$ are the summation with respect to the permutations $\sigma_1$ for $\{1,2,\ldots,m_1\}$ and $\sigma_2$ for 
$\{1,2,\ldots,m_2\}$, respectively.
\end{prop}

\begin{figure}
\centering
\includegraphics[clip, scale=0.6]{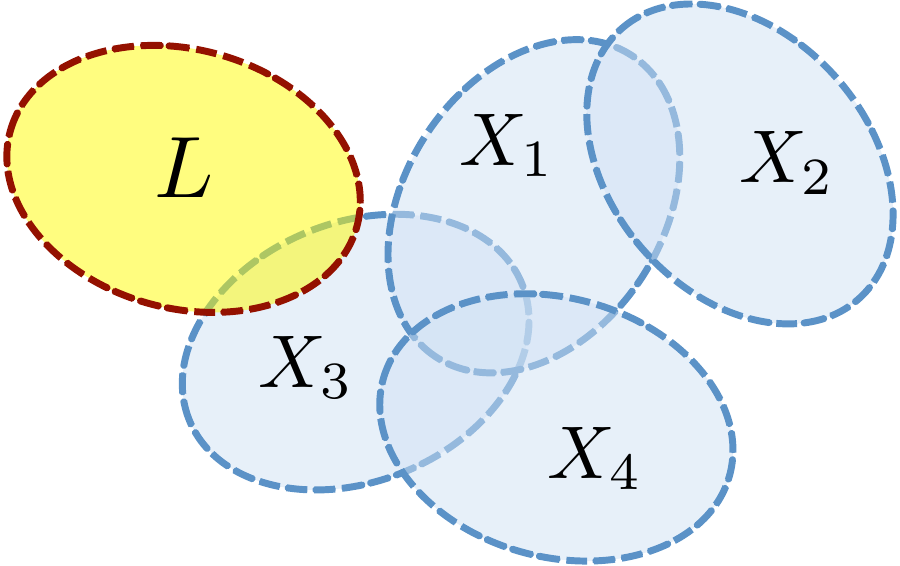}
\caption{$N_{\ed_s | w_L}$ is defined by a number of subsets in $w$ that have overlap with $\ed_s$.
When $w_L=\{L,\ed_1,\ed_2,\ed_3,\ed_4\}$ is given as above, we have $N_{\ed_1 | w_L}=3$, $N_{\ed_2 | w_L}=3$, $N_{\ed_3 | w_L}=2$ and $N_{\ed_4| w_L}=1$.}
\label{fig:Number_overlap}
\end{figure}

We then obtain an upper bound of $|\Der_{w_2} \tilde{\Der}_{w_1} \tr \bigl( f_\ed \Phi_{\vec{a},\vec{b}}   \bigr)  |$ for $\vec{a}=\vec{0}$ and $\vec{b}=\vec{0}$.
From the explicit form given in Proposition~\ref{prop:explicit_form_derivation_CH}, we have
\begin{align}
&\left| \Der_{w_2} \tilde{\Der}_{w_1} \tr \left( f_\ed \Phi_{\vec{a},\vec{b}}    \right) \Bigl|_{\vec{a}=0,\vec{b}=0} \right|  \notag \\
\le
&\beta^{m_2} |\tau|^{m_1}\max_{\sigma_1,\sigma_2} \left |
\copyave{m+1}{
f_\ed^{(0)} f_{\ed_{1,\sigma_1(1)}}^{(1)} f_{\ed_{1,\sigma_1(2)}}^{(2)} \cdots f_{\ed_{1,\sigma_1(m_1)}}^{(m_1)} 
h_{\ed_{2,\sigma_2(1)}}^{(m_1+1)} h_{\ed_{2,\sigma_2(2)}}^{(m_1+2)} \cdots h_{\ed_{2,\sigma_2(m_2)}}^{(m)} 
}\right|. \label{Der_w_1_w_2_bound_up}
\end{align}
In order to bound the right-hand side of \eqref{Der_w_1_w_2_bound_up}, we utilize the following proposition (see \ref{proof_prop:bound_on_Der} for the proof):
\begin{prop} \label{prop:bound_on_Der}
Let $O_0\in \ban(\mathcal{H})$ be an operator supported on a subset $L\subseteq V$ such that $\tr(O_0)=0$. 
We then consider arbitrary $m$ operators $\{O_s\}_{s=1}^{m}$ ($O_s\in \ban(\mathcal{H})$) which are supported on $w:=\{\ed_s\}_{s=1}^{m}$ respectively and satisfy $\tr(O_s)=0$ for $s=1,2\ldots,m$. For each of $\{O_s\}_{s=0}^m$, we define $O_s^{(s)}$ as in Eq.~\eqref{extended_operator_prop}.
We then prove 
\begin{align}
\left|\copyave{m+1}{ O_{0}^{(0)} O_{1}^{(1)} O_{2}^{(2)}\cdots O_{m}^{(m)} }  \right| \le  \|O_{0}\|   \prod_{s=1}^{m} 
2 N_{\ed_s | w_L} \|O_{s}\| .
\label{ineq_Der_norm_overlap_form}
\end{align}
where $w_L:= \{L,\ed_1,\ed_2,\ldots ,\ed_m\}$ and   
$N_{\ed_s | w_L}$ is a number of subsets in $w_L$ that have overlap with $\ed_s$ (Fig.~\ref{fig:Number_overlap}):
\begin{align}
N_{\ed_s | w_L} = \# \{\ed \in w_L|  \ed\neq \ed_s ,  \ed \cap \ed_s \neq \emptyset  \}.
\end{align}
\end{prop}

\textit{Remark.} We note that the following simple estimation cannot be used.
Because Eq.~\eqref{extended_operator_prop} gives $\| O^{(s)}\| \le 2s \|O\|$, we immediately obtain 
\begin{align}
 \left |\copyave{m+1}{
f_\ed^{(0)} f_{\ed_{1,1}}^{(1)}\cdots f_{\ed_{1,m_1}}^{(m_1)} h_{\ed_{2,1}}^{(m_1+1)} \cdots h_{\ed_{2,m_2}}^{(m)}} \right|  
  \le  2^{m} m_1! m_2! \|f_\ed\| \prod_{s=1}^{m_1}\| f_{\ed_{1,s}}\|  \prod_{s'=1}^{m_2} \| h_{\ed_{2,s'}}\| . 
 \end{align}
However, this estimation is too loose and cannot ensure the convergence of Eq.~\eqref{DMCE_Cluster_decomp_CH_ineq}. 
We thus need more refined analysis, and Proposition~\ref{prop:bound_on_Der} plays a crucial role in proving the convergence of the cluster expansion.

{~}\\

By applying the proposition~\ref{prop:bound_on_Der} to the upper bound~\eqref{Der_w_1_w_2_bound_up}, we have 
\begin{align}
&\left| \Der_{w_2} \tilde{\Der}_{w_1} \tr \left( f_\ed \Phi_{\vec{a},\vec{b}}    \right) \Bigl|_{\vec{a}=0,\vec{b}=0} \right|
\le (2\beta)^{m_2}(2|\tau|)^{m_1} \|f_\ed\| \prod_{s_1=1}^{m_1} N_{\ed_{1,s_1} | w_\ed}  \| f_{\ed_{1,s_1}}\|       
\prod_{s_2=1}^{m_2}  N_{\ed_{2,s_2} | w_\ed}  \|h_{\ed_{2,s_2}}\| ,
\end{align}
where $w_\ed=\{\ed, \ed_{1,1},\ldots, \ed_{1,m_1},\ed_{2,1},\ldots, \ed_{2,m_2}\}$.
From the above inequality, the equation~\eqref{DMCE_Cluster_decomp_CH_ineq} is bounded from above by 
\begin{align}
\log \left[\tr \left(e^{-\tau \Fp_{\vec{1}}} \rho_{\vec{1}}\right) \right]   +\tau \tr (\Fp\rho_{\vec{1}})  
=&\sum_{\ed \in E_k}
\sum_{m_1=1}^{\infty} \sum_{m_2=0}^{\infty}  \sum_{\substack{w_1\in \Cl_{m_1}, w_2 \in \Cl_{m_2}\\ w_1\oplus w_2 \in \Gc_{m_1+m_2}^X}}\frac{-\tau n_{w_1}n_{w_2} }{(m_1+1)!m_2!} \Der_{w_2} \tilde{\Der}_{w_1}  \tr \left( f_\ed \Phi_{\vec{a},\vec{b}}   \right)\Bigl|_{\vec{a}=\vec{0},\vec{b}=\vec{0}} \notag \\
\le& \sum_{\ed \in E_k} 
\sum_{m_1=1}^{\infty} \sum_{m_2=0}^{\infty}  \sum_{\substack{w_1\in \Cl_{m_1}, w_2 \in \Cl_{m_2}\\ w_1\oplus w_2 \in \Gc_{m_1+m_2}^X}}\frac{|\tau|  (2\beta)^{m_2}(2|\tau|)^{m_1} n_{w_1}n_{w_2} }{(m_1+1)!m_2!} \Psi_{w_X}^{(m_1,m_2)}   
\label{expect_cluster_m_2_equal_to_0_CH_0}
\end{align}
with $\Psi_{w_X}^{(m_1,m_2)}:=\|f_\ed\| \prod_{s_1=1}^{m_1} N_{\ed_{1,s_1} | w_\ed}  \| f_{\ed_{1,s_1}}\|\prod_{s_2=1}^{m_2}  N_{\ed_{2,s_2} | w_\ed}  \|h_{\ed_{2,s_2}}\|$, 
where we use Eq.~\eqref{der_b_e:phi_a_b} in the first equation.
For a fixed $\{w, m_1,m_2\}$, the number of combinations for $w_1$ and $w_2$ satisfying $w_1\oplus w_2=w$ is smaller than $\binom{m_1+m_2}{m_1}$.  
Also, we have $n_w \ge n_{w_1} n_{w_2}$, and hence the inequality~\eqref{expect_cluster_m_2_equal_to_0_CH_0} further reduces to
\begin{align}
\log \left[\tr \left(e^{-\tau \Fp_{\vec{1}}} \rho_{\vec{1}}\right) \right]   +\tau \tr (\Fp\rho_{\vec{1}})  
&\le \sum_{\ed \in E_k} \sum_{m=1}^\infty \sum_{m_1=1}^m\sum_{w\in \Gc_m^X} \binom{m}{m_1}\frac{|\tau|  (2\beta)^{m-m_1}(2|\tau|)^{m_1} n_w }{(m_1+1)!(m-m_1)!}  \Psi_{w_X}^{(m_1,m-m_1)}  \notag \\
&\le \sum_{\ed \in E_k}  \sum_{m=1}^\infty 2^{m-1}|\tau|  \sum_{m_1=1}^m\binom{m}{m_1}(2\beta)^{m-m_1}(2|\tau|)^{m_1}  \sum_{w\in \Gc_m^X} \frac{n_w}{m!}  \Psi_{w_X}^{(m_1,m-m_1)} 
 , \label{expect_cluster_m_2_equal_to_0_CH}
\end{align}
where in the first inequality we set $m_1+m_2=m$, and in the second inequality, we use $\frac{1}{(m_1+1)!(m-m_1)!}\le \frac{2^{m-1}}{m!} $.

%

In order to estimate the convergence rate of the expansion in \eqref{expect_cluster_m_2_equal_to_0_CH}, 
we prove the following proposition (see \ref{proof_prop: convergence of cluster expansion} for the proof):
\begin{prop} \label{prop: convergence of cluster expansion}
Let $\{o_\ed\}_{\ed\in E_k}$ be arbitrary operators such that 
\begin{align}
\sum_{\ed: \ed \ni v}\| o_{\ed}\|\le g     \for \forall v\in V \label{cond_for_o_ed}
\end{align}
Then, for an arbitrary subset $L$, we obtain
\begin{align}
\sum_{w \in \Gc^L_{m}} \frac{n_{w} }{m!} \prod_{s=1}^{m} N_{\ed_s | w_L}  \| o_{\ed_{s}}\|    \le \frac{1}{2}  e^{|L|/k} (2e^{3} gk)^m .
 \end{align}
\end{prop}

By using Proposition~\ref{prop: convergence of cluster expansion} with $L=\ed$, we can derive an upper bound of 
 \begin{align}
 \sum_{w\in \Gc_m^X} \frac{n_w}{m!}  \Psi_{w_X}^{(m_1,m-m_1)} \le \frac{e}{2} (2e^3gk)^{m} \|f_X\|,
\end{align}
which yields,
 \begin{align}
\sum_{m_1=1}^m\binom{m}{m_1}(2\beta)^{m-m_1}(2|\tau|)^{m_1}  \sum_{w\in \Gc_m^X} \frac{n_w}{m!}  \Psi_{w_X}^{(m_1,m-m_1)} 
 \le& \sum_{m_1=1}^{m} \binom{m}{m_1} (2\beta)^{m-m_1}(2|\tau|)^{m_1}  \frac{e}{2} (2e^3gk)^{m} \|f_X\| \notag \\
 =& \frac{e}{2}\|f_X\| (4e^3gk)^{m}[(\beta+|\tau|)^m - \beta^m] .
\end{align}
This reduces the inequality~\eqref{expect_cluster_m_2_equal_to_0_CH} to 
\begin{align}
&\log \left[\tr \left(e^{-\tau \Fp_{\vec{1}}} \rho_{\vec{1}}\right) \right]   +\tau \tr (\Fp\rho_{\vec{1}})  \notag \\
\le&\sum_{\ed\in E_k}  \|f_\ed\|  \sum_{m=1}^{\infty}2^{m-1}|\tau|    \frac{e}{2}  (4e^3gk)^{m}[(\beta+|\tau|)^m - \beta^m]  , \notag \\
\le& \frac{e |\tau|}{4} \bar{\Fp} \left( \frac{(\beta+|\tau|)/\beta_c}{1-(\beta+|\tau|)/\beta_c} - \frac{\beta/\beta_c}{1-\beta/\beta_c}\right) 
 \le \frac{\tau^2 /\beta_c}{1-(\beta+|\tau|)/\beta_c} \bar{\Fp}  =  \frac{\tau^2 \bar{\Fp}}{\beta_c-\beta-|\tau|}   ,
\end{align}
where we use the notation of $\beta_c=1/(8e^3gk)$ and the notation of $\bar{\Fp}$ as $\bar{\Fp} := \sum_{|\ed |\le k}\| f_\ed \|$.
This completes the proof of Theorem~\ref{main_theorem_CH_inequality}. $\square$

\section{acknowledgments}
The work of T. K. was supported by the RIKEN Center for AIP and JSPS KAKENHI Grant No. 18K13475.
K.S. was supported by JSPS Grants-in-Aid for Scientific Research (Grants No. JP16H02211, No. JP19H05791, and No. JP19H05603).

\appendix

\section{Proof of Proposition~\ref{prop: connceted_cluster_CH_ineq}} \label{proof_prop: connceted_cluster_CH_ineq}

For the proof, we first rewrite
\begin{align}
\Der_{w_2} \tilde{\Der}_{w_1} \log \left[\tr \left(e^{-\tau \Fp_{\vec{b}}} \rho_{\vec{a}}\right) \right] \Bigl|_{\vec{a}=\vec{0},\vec{b}=\vec{0}} =
\Der_{w_2} \tilde{\Der}_{w_1} \log \left[\tr \left(e^{-\tau \Fp_{\vec{b}_{w_1}}} \rho_{\vec{a}_{w_2}}\right) \right] \Bigl|_{\vec{a}_{w_2}=\vec{0},\vec{b}_{w_1}=\vec{0}},
\end{align}
where we define $\vec{a}_{w}$ as a parameter vector such that only the elements in $w$ are non-zero, namely
\begin{align}
( \vec{a}_{w} )_{\ed} \begin{cases}  \neq 0 \for  \ed \in  w,\\ =0 \for  \ed \notin  w . \end{cases} \label{def:a_w_para}
\end{align}
An element of $a_{\ed}$ in $\vec{a}$ is denoted by $( \vec{a})_{\ed}$.
In the same way, we define $\vec{b}_{w}$.

We now need to prove
\begin{align}
\Der_{w_2} \tilde{\Der}_{w_1} \log \left[\tr \left(e^{-\tau \Fp_{\vec{b}_{w_1}}} \rho_{\vec{a}_{w_2}}\right) \right] \Bigl|_{\vec{a}_{w_2}=\vec{0},\vec{b}_{w_1}=\vec{0}} =0 \label{ineq: prop: connceted_cluster_CH_ineq}
\end{align}
for $w_1 \oplus w_2 \notin \Gc_{|w_1|+|w_2|}$. 
The unconnected condition of the cluster $w_1 \oplus w_2$ implies the existence of the decomposition of
\begin{align}
w_1 \oplus w_2 = (\tilde{w}_1 \oplus \tilde{w}_2) \oplus (\tilde{\tilde{w}}_1 \oplus \tilde{\tilde{w}}_2) ,\quad V_{\tilde{w}_1 \oplus \tilde{w}_2} \cap V_{\tilde{\tilde{w}}_1 \oplus \tilde{\tilde{w}}_2}=\emptyset, \label{decomp_w1_w2_unconnecte}
\end{align}
where $w_1=\tilde{w}_1\oplus \tilde{\tilde{w}}_1$ and $w_2=\tilde{w}_2\oplus \tilde{\tilde{w}}_2$ with $|\tilde{w}_1 \oplus \tilde{w}_2| >0$ and $|\tilde{\tilde{w}}_1 \oplus \tilde{\tilde{w}}_2|>0$. 
Then, the operator $\Fp_{\vec{b}_{w_1}}=\Fp_{\vec{b}_{\tilde{w}_1\oplus \tilde{\tilde{w}}_1}}$ can be decomposed as 
\begin{align}
&\Fp_{\vec{b}_{\tilde{w}_1\oplus \tilde{\tilde{w}}_1}} =  \Fp_{\vec{b}_{\tilde{w}_1}}  + \Fp_{\vec{b}_{\tilde{\tilde{w}}_1}} ,
\end{align}
which yields
\begin{align}
e^{-\tau \Fp_{\vec{b}_{w_1}}} =  e^{-\tau \Fp_{\vec{b}_{\tilde{w}_1}} }  \otimes e^{-\tau  \Fp_{\vec{b}_{\tilde{\tilde{w}}_1}} }  \label{decomp_omega_b}
\end{align}
because of $V_{\tilde{w}_1} \cap V_{\tilde{\tilde{w}}_1}=\emptyset$.
Notice that the operators $ \Fp_{\vec{b}_{\tilde{w}_1}} $ and $\Fp_{\vec{b}_{\tilde{\tilde{w}}_1}}$ are supported on the subsets $V_{\tilde{w}_1}$ and $V_{\tilde{\tilde{w}}_1}$, respectively.
Also, the density matrix $\rho_{\vec{a}_{w_2}}=\rho_{\vec{a}_{\tilde{w}_2\oplus \tilde{\tilde{w}}_2}}$ is now defined by using the Hamiltonian 
\begin{align}
&H_{\vec{a}_{w_2}}= H_{\vec{a}_{\tilde{w}_2\oplus \tilde{\tilde{w}}_2}} =  H_{\vec{a}_{\tilde{w}_2}}  + H_{\vec{a}_{\tilde{\tilde{w}}_2}} ,
\end{align}
and hence $e^{-\beta H_{\vec{a}_{w_2}}} =e^{-\beta H_{\vec{a}_{\tilde{w}_2}}} \otimes e^{-\beta H_{\vec{a}_{\tilde{\tilde{w}}_2}}} $. 
Therefore, from Eqs.~\eqref{decomp_w1_w2_unconnecte} and \eqref{decomp_omega_b}, we obtain
\begin{align}
&\tr \left(e^{-\tau \Fp_{\vec{b}_{w_1}}} \rho_{\vec{a}_{w_2}} \right)=  \tr \left( e^{-\tau \Fp_{\vec{b}_{\tilde{w}_1}} } \rho_{\vec{a}_{\tilde{w}_2}}  \right) 
\tr \left(e^{-\tau  \Fp_{\vec{b}_{\tilde{\tilde{w}}_1}} } \rho_{\vec{a}_{\tilde{\tilde{w}}_2}}  \right)  .\label{decomp_rho_a_w_2}
\end{align}

The equation~\eqref{decomp_rho_a_w_2} implies
\begin{align}
\Der_{w_2} \tilde{\Der}_{w_1} \log \left[\tr \left(e^{-\tau \Fp_{\vec{b}_{w_1}}} \rho_{\vec{a}_{w_2}}\right) \right] =&\Der_{\tilde{w}_2} \Der_{\tilde{\tilde{w}}_2} \tilde{\Der}_{\tilde{w}_1}\tilde{\Der}_{\tilde{\tilde{w}}_1}  
 \left(\log \left[\tr \left( e^{-\tau \Fp_{\vec{b}_{\tilde{w}_1}} } \rho_{\vec{a}_{\tilde{w}_2}}  \right) \right] 
+\log \left[\tr \left(e^{-\tau  \Fp_{\vec{b}_{\tilde{\tilde{w}}_1}} } \rho_{\vec{a}_{\tilde{\tilde{w}}_2}}  \right) \right]  \right). \label{derivation_decomp_w2_w1_1}
\end{align}
Finally, because of 
 \begin{align} 
&\Der_{\tilde{\tilde{w}}_2} \tilde{\Der}_{\tilde{\tilde{w}}_1}  \log \left[\tr \left( e^{-\tau \Fp_{\vec{b}_{\tilde{w}_1}} } \rho_{\vec{a}_{\tilde{w}_2}}  \right) \right]    =0 ,
\quad \Der_{\tilde{w}_2} \tilde{\Der}_{\tilde{w}_1} \log \left[\tr \left(e^{-\tau  \Fp_{\vec{b}_{\tilde{\tilde{w}}_1}} } \rho_{\vec{a}_{\tilde{\tilde{w}}_2}}  \right) \right]  =0
\end{align}
for $|\tilde{w}_1 \oplus \tilde{w}_2| >0$ and $|\tilde{\tilde{w}}_1 \oplus \tilde{\tilde{w}}_2|>0$, Eq.~\eqref{derivation_decomp_w2_w1_1} reduces to
\begin{align}
&\Der_{w_2} \tilde{\Der}_{w_1} \log \left[\tr \left(e^{-\tau \Fp_{\vec{b}_{w_1}}} \rho_{\vec{a}_{w_2}}\right) \right]=0.
\end{align}
This gives the equation~\eqref{ineq: prop: connceted_cluster_CH_ineq} and completes the proof of Proposition~\ref{prop: connceted_cluster_CH_ineq}. 
$\square$

%
%
%

\section{Proof of Proposition~\ref{prop:explicit_form_derivation_CH}} \label{proof_prop:explicit_form_derivation_CH}

We here show the proof of Proposition~\ref{prop:explicit_form_derivation_CH}, which gives the explicit form of $\Der_{w_2} \tilde{\Der}_{w_1} \tr \bigl( f_\ed \Phi_{\vec{a},\vec{b}}   \bigr)\bigl |_{\vec{a}=\vec{0},\vec{b}=\vec{0}}$.
For the proof, we first consider the Taylor expansion with respect to $\beta$ and $\tau$:
\begin{align}
\tr \left( f_\ed \Phi_{\vec{a},\vec{b}}   \right) =
 \sum_{m_1=0}^{\infty} \sum_{m_2=0}^{\infty}\frac{\tau^{m_1}}{m_1!}\frac{\beta^{m_2}}{m_2!}
\frac{\partial^{m_2}}{\partial \beta^{m_2}}\frac{\partial^{m_1}}{\partial \tau^{m_1}}\tr \left( f_\ed \Phi_{\vec{a},\vec{b}}   \right)  \biggl |_{\beta=0,\tau=0} \label{taylor_exp_omega_ed_phi}
\end{align}



The following lemma gives the explicit form of the derivatives:
\begin{lemma} \label{thm:express_Der_operator}
By using the notation~\eqref{extended_operator_prop}, we obtain derivatives of $\log \left[\tr \left(e^{-\tau \Fp_{\vec{b}}} e^{-\beta H_{\vec{a}}}\right) \right]$ 
with respect to $\beta$ and $\tau$ as 
\begin{align}
\frac{\partial^{m_2}}{\partial \beta^{m_2}}\frac{\partial^{m_1}}{\partial \tau^{m_1}}\tr \left( f_\ed \Phi_{\vec{a},\vec{b}}   \right)   
= (-1)^m  \tr_{\mathcal{H}^{\otimes m+1}}  \left(  f_\ed^{(0)}\Fp_{\vec{b}}^{(1)}\cdots \Fp_{\vec{b}}^{(m_1)} \Phi_{\vec{a},\vec{b}}^{\otimes m+1} H_{\vec{a}}^{(m_1+1)} H_{\vec{a}}^{(m_1+2)} \cdots H_{\vec{a}}^{(m)} \right), \label{tau_beta_derivative}
\end{align}
where $m=m_1+m_2$.
This yields for $\beta=\tau=0$ 
 \begin{align}
\frac{\partial^{m_2}}{\partial \beta^{m_2}}\frac{\partial^{m_1}}{\partial \tau^{m_1}}\tr \left( f_\ed \Phi_{\vec{a},\vec{b}}   \right) \biggl |_{\beta=0,\tau=0}
= (-1)^m\copyave{m+1}{f_\ed^{(0)}\Fp_{\vec{b}}^{(1)}\cdots \Fp_{\vec{b}}^{(m_1)}  H_{\vec{a}}^{(m_1+1)} H_{\vec{a}}^{(m_1+2)} \cdots H_{\vec{a}}^{(m)} },
\end{align} 
where we use the notation~\eqref{def:copyave}, $\Phi_{\vec{a},\vec{b}}= \hat{1}/D_\mathcal{H}$ for $\beta=0$ and $\tau=0$ (see the definition~\eqref{def:phi_a_b}).
\end{lemma} 

By taking $\Der_{w_2} \tilde{\Der}_{w_1}$ with $|w_1|=m_1$ and $|w_2|=m_2$, only the $m_1$th order terms of $\tau$ and $m_2$th order terms of $\beta$ survives in Eq.~\eqref{taylor_exp_omega_ed_phi}, respectively. 
Hence, we have
\begin{align}
&\Der_{w_2} \tilde{\Der}_{w_1} \tr \left( f_\ed \Phi_{\vec{a},\vec{b}}   \right)   \bigl|_{\vec{a}=0,\vec{b}=0}  \notag \\
=& \frac{\tau^{m_1}}{m_1!} \frac{\beta^{m_2}}{m_2!}\Der_{w_2} \tilde{\Der}_{w_1} \left( \frac{\partial^{m_2}}{\partial \beta^{m_2}}\frac{\partial^{m_1}}{\partial \tau^{m_1}}\tr \left( f_\ed \Phi_{\vec{a},\vec{b}}   \right) \biggl |_{\beta=0,\tau=0} \right) \notag \\
=&\frac{(-\tau)^{m_1}(-\beta)^{m_2}}{m_1! m_2!} \Der_{w_2} \tilde{\Der}_{w_1}\copyave{m+1}{f_\ed^{(0)}\Fp_{\vec{b}}^{(1)}\cdots \Fp_{\vec{b}}^{(m_1)}  H_{\vec{a}}^{(m_1+1)} H_{\vec{a}}^{(m_1+2)} \cdots H_{\vec{a}}^{(m)} } \biggl|_{\vec{a}=0,\vec{b}=0}, \label{prop_der_w1_w2_first_eq}
\end{align}
where the second equation comes from Lemma~\ref{thm:express_Der_operator}.

By using the permutations of $\sigma_1$ and $\sigma_2$, we obtain
\begin{align}
&\tilde{\Der}_{w_1} \Fp_{\vec{b}}^{(1)}\Fp_{\vec{b}}^{(2)}\cdots \Fp_{\vec{b}}^{(m_1)} \bigl|_{\vec{b}=0} 
=\sum_{\sigma_1} f_{\ed_{1,\sigma_1(1)}}^{(1)} f_{\ed_{1,\sigma_1(2)}}^{(2)} \cdots f_{\ed_{1,\sigma_1(m_1)}}^{(m_1)}  , \notag \\
&\Der_{w_2} H_{\vec{a}}^{(m_1+1)} H_{\vec{a}}^{(m_1+2)} \cdots H_{\vec{a}}^{(m)}  \bigl|_{\vec{a}=0}
= \sum_{\sigma_2} h_{\ed_{2,\sigma_2(1)}}^{(m_1+1)} h_{\ed_{2,\sigma_2(2)}}^{(m_1+2)} \cdots h_{\ed_{2,\sigma_2(m_2)}}^{(m)}  . \label{prop_der_w1_w2_second_eq}
\end{align}
By combining Eqs.~\eqref{prop_der_w1_w2_first_eq} and \eqref{prop_der_w1_w2_second_eq}, we obtain Eq.~\eqref{simple_expression_der_CH_ineq}.
This completes the proof of Proposition~\ref{prop:explicit_form_derivation_CH}.  $\square$

{~}

\textit{Proof of Lemma~\ref{thm:express_Der_operator}.}
For the proof, we relies on the induction method.
For $m_1+m_2=1$, we can obtain Eq.~\eqref{tau_beta_derivative} because of
\begin{align}
&\frac{\partial}{\partial \beta} \tr \left( f_\ed \Phi_{\vec{a},\vec{b}}   \right) 
= \tr \left( f_\ed \Phi_{\vec{a},\vec{b}} H_{\vec{a}}   \right) -  \tr \left( f_\ed \Phi_{\vec{a},\vec{b}}\right) \cdot \tr \left(\Phi_{\vec{a},\vec{b}} H_{\vec{a}}   \right) 
= \tr_{\mathcal{H}^{\otimes 2}} \left( f_\ed^{(0)} \Phi_{\vec{a},\vec{b}}^{\otimes 2} H_{\vec{a}}^{(1)}  \right)
\end{align}
and
\begin{align}
&\frac{\partial}{\partial \tau} \tr \left( f_\ed \Phi_{\vec{a},\vec{b}}   \right) 
= \tr \left( f_\ed  \Fp_{\vec{b}}  \Phi_{\vec{a},\vec{b}}    \right) -  \tr \left( f_\ed \Phi_{\vec{a},\vec{b}}\right) \cdot \tr \left(\Fp_{\vec{b}}  \Phi_{\vec{a},\vec{b}} \right) 
= \tr_{\mathcal{H}^{\otimes 2}} \left( f_\ed^{(0)} \Fp_{\vec{b}}^{(1)} \Phi_{\vec{a},\vec{b}}^{\otimes 2}  \right)
\end{align}

We then assume Eq.~\eqref{tau_beta_derivative} for $m_1+m_2=m-1$ and prove the case of $m_1+m_2=m$.
We first consider the case of $\frac{\partial^{m_2+1}}{\partial \beta^{m_2+1}}\frac{\partial^{m_1}}{\partial \tau^{m_1}}$ with $m_1+m_2=m-1$.
The assumption for $m_1+m_2=m-1$ gives 
\begin{align}
\frac{\partial^{m_2+1}}{\partial \beta^{m_2+1}}\frac{\partial^{m_1}}{\partial \tau^{m_1}} \tr \left( f_\ed \Phi_{\vec{a},\vec{b}}   \right) 
=&\frac{\partial}{\partial \beta}  (-1)^{m-1}  \tr_{\mathcal{H}^{\otimes m+1}}  \left( f_\ed^{(0)}\Fp_{\vec{b}}^{(1)}\cdots \Fp_{\vec{b}}^{(m_1)} \Phi_{\vec{a},\vec{b}}^{\otimes m+1} H_{\vec{a}}^{(m_1+1)} H_{\vec{a}}^{(m_1+2)} \cdots H_{\vec{a}}^{(m-1)} \right), \label{tau_beta_derivative_m}
\end{align}
Then, we have
\begin{align}
\frac{\partial }{\partial \beta} \Phi_{\vec{a},\vec{b}}^{\otimes m+1}  
&= \frac{\partial }{\partial \beta} 
\left( \frac{e^{-\tau \Fp_{\vec{b}}} e^{-\beta H_{\vec{a}}}} {\tr (e^{-\tau \Fp_{\vec{b}}} e^{-\beta H_{\vec{a}}})}\right)^{\otimes m} \notag \\
& = -\Phi_{\vec{a},\vec{b}}^{\otimes m+1} \sum_{s=1}^{m} H_{\vec{a},\mathcal{H}_s} 
+m \left(e^{-\tau \Fp_{\vec{b}}} e^{-\beta H_{\vec{a}}} \right)^{\otimes m} 
\frac{\tr (e^{-\tau \Fp_{\vec{b}}} e^{-\beta H_{\vec{a}}} H_{\vec{a}} )}{[\tr (e^{-\tau \Fp_{\vec{b}}} e^{-\beta H_{\vec{a}}})]^{m+1}}  \notag \\
&= -\Phi_{\vec{a},\vec{b}}^{\otimes m+1} \sum_{s=1}^{m} H_{\vec{a},\mathcal{H}_s} 
+\tr_{\mathcal{H}_{m+1}} \left(  m \Phi_{\vec{a},\vec{b}}^{\otimes m+1} H_{\vec{a},\mathcal{H}_{m+1}}\right) =-\tr_{\mathcal{H}_{m+1}}\left( \Phi_{\vec{a},\vec{b}}^{\otimes m+1} H_{\vec{a}}^{(m)}\right) \label{tau_beta_derivative_m2}
\end{align}
By combining Eqs.~\eqref{tau_beta_derivative_m} and \eqref{tau_beta_derivative_m2}, we obtain
\begin{align}
&\frac{\partial^{m_2+1}}{\partial \beta^{m_2+1}}\frac{\partial^{m_1}}{\partial \tau^{m_1}} \tr \left( f_\ed \Phi_{\vec{a},\vec{b}}   \right)   \notag  \\
=& (-1)^{m}  \tr_{\mathcal{H}^{\otimes m+1}}  \left( f_\ed^{(0)} \Fp_{\vec{b}}^{(1)}\cdots \Fp_{\vec{b}}^{(m_1)} \Phi_{\vec{a},\vec{b}}^{\otimes m+1} H_{\vec{a}}^{(m)} H_{\vec{a}}^{(m_1+1)} H_{\vec{a}}^{(m_1+2)} \cdots H_{\vec{a}}^{(m-1)} \right). 
\end{align}
We thus obtain Eq.~\eqref{tau_beta_derivative} by using $[H_{\vec{a}}^{(s)},H_{\vec{a}}^{(s')}]=0$ for $ \forall s,s'$. 
The same analysis can be applied to the case of $\frac{\partial^{m_2}}{\partial \beta^{m_2}}\frac{\partial^{m_1+1}}{\partial \tau^{m_1+1}}$.
This completes the proof of  Eq.~\eqref{tau_beta_derivative} for $m_1+m_2=m$. $\square$

%
%
%
%
%

\section{Proof of Proposition~\ref{prop:bound_on_Der}} \label{proof_prop:bound_on_Der}

For the proof, we first notice that from Eq.~\eqref{def:copyave},
$\copyave{m+1}{O_{0}^{(0)}  O_{1}^{(1)} O_{2}^{(2)}\cdots O_{m}^{(m)}}$ consists of a summation of multiplications as follows:
\begin{align}
\frac{\tr (O_{u_1})}{D_{\mathcal{H}}}\cdot \frac{\tr (O_{u_2})}{D_{\mathcal{H}}}  \cdots \frac{\tr (O_{u_q})}{D_{\mathcal{H}}} ,
\label{multi_decomp_O_u}
\end{align}
where each of $\{u_j\}_{j=1}^q$ is an integer subset in $\{0, 1,2,3,\ldots,m\}$ with $|u_j|\ge 2$ and $u_1\oplus u_2\oplus \cdots \oplus u_q=\{0,1,2,3,\ldots,m\} $; also for $u=\{i_1,i_2,\ldots, i_{|u|}\}$, we define
$O_u:=O_{l_1}O_{i_2} \cdots O_{i_{|u|}}$ with $0\le i_1<i_2 < \cdots < i_{|u|} \le m$.
We notice that each of the sets $\{u_j\}_{j=1}^q$ in \eqref{multi_decomp_O_u} is irreducible in the sense that $\{\ed_{i_1},\ed_{i_2},\ldots,\ed_{i_{|u|}}\} \in \Gc_{|u|}$ ($u=\{i_1,i_2,\ldots,i_{|u|}\}$).

We then obtain the following decomposition:
\begin{align}
\copyave{m+1}{O_{0}^{(0)}  O_{1}^{(1)} O_{2}^{(2)}\cdots O_{m}^{(m)}} = \sum_{q=1}^{n/2}\sum_{u_1,u_2,\ldots,u_q} \mathcal{N}_{u_1,u_2,\ldots,u_q}\frac{\tr (O_{u_1})}{D_{\mathcal{H}}}\cdot \frac{\tr (O_{u_2})}{D_{\mathcal{H}}}  \cdots \frac{\tr (O_{u_q})}{D_{\mathcal{H}}} ,
 \label{summation_of_all_possibility_u}
\end{align}
where $\mathcal{N}_{u_1,u_2,\ldots,u_q} \in \mathbb{Z}$ is a non-trivial coefficient which can be calculated from~\eqref{extended_operator_prop}.

\begin{figure}
\centering
\includegraphics[clip, scale=0.6]{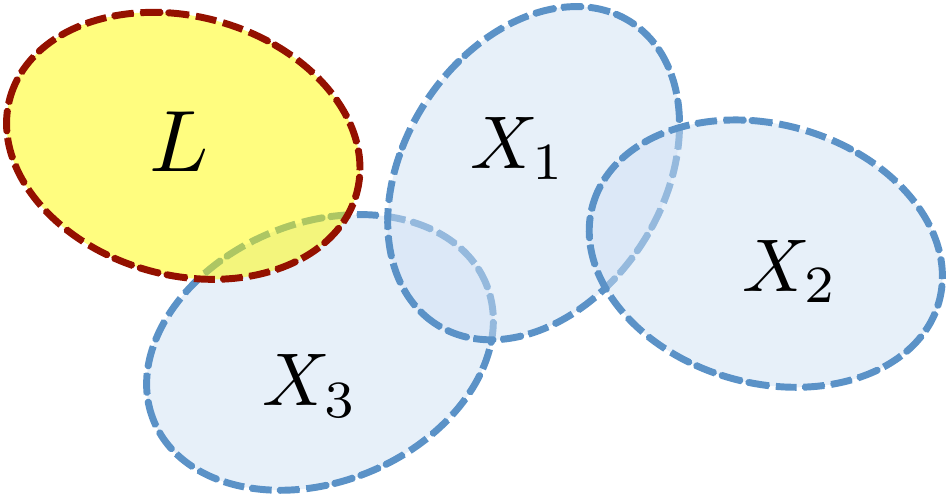}
\caption{The definition of the subsets $\{L,X_1,X_2,X_3\}$ which support the operators $\{O_0,O_1,O_2,O_3\}$, respectively.}
\label{fig:Prop_3}
\end{figure}

For example, let us consider the case of $m=4$ as shown in Fig.~\ref{fig:Prop_3}.
Because of $\tr (O_s)=0$ for $s=0,1,2,3$, we have 
\begin{align}
\copyave{4}{O_{0}^{(0)}  O_{1}^{(1)} O_{2}^{(2)}\cdots O_{3}^{(3)}} = &
 \frac{\tr (O_{0} O_{1}O_{2}O_{3}) }{D_{\mathcal{H}}}\notag \\
&  - \frac{\tr (O_{0} O_{1}) }{D_{\mathcal{H}}} \frac{\tr (O_{2}O_{3}) }{D_{\mathcal{H}}}
 -\frac{\tr (O_{0} O_{2}) }{D_{\mathcal{H}}} \frac{\tr (O_{1}O_{3}) }{D_{\mathcal{H}}}  -\frac{\tr (O_{0} O_{3}) }{D_{\mathcal{H}}} \frac{\tr (O_{1}O_{2}) }{D_{\mathcal{H}}}
\end{align}
Then, due to $L\cap X_1=L\cap X_2=\emptyset$ from Fig.~\ref{fig:Prop_3}, only the sets of $\{0,1,2,3\}$, $\{0,3\}$ and $\{1,2\}$ are irreducible.
We thus obtain
\begin{align}
\copyave{4}{O_{0}^{(0)}  O_{1}^{(1)} O_{2}^{(2)}\cdots O_{3}^{(3)}}  = &
 \frac{\tr (O_{0} O_{1}O_{2}O_{3}) }{D_{\mathcal{H}}} -\frac{\tr (O_{0} O_{3}) }{D_{\mathcal{H}}} \frac{\tr (O_{1}O_{2}) }{D_{\mathcal{H}}}.
\end{align}
This yields $\mathcal{N}_{\{0,1,2,3\}}=1$ and $\mathcal{N}_{\{0,3\},\{1,2\}}=-1$. 

Our task is now to estimate the upper bound of
\begin{align}
\mathcal{N}\left ( O_{0}^{(0)}  O_{1}^{(1)} O_{2}^{(2)}\cdots O_{m}^{(m)}  \right) :=\sum_{q=1}^{n/2}\sum_{u_1,u_2,\ldots,u_q} |\mathcal{N}_{u_1,u_2,\ldots,u_q}|  .
\end{align}
Because of 
\begin{align}
\frac{\tr (O_{u_1})}{D_{\mathcal{H}}}\cdot \frac{\tr (O_{u_2})}{D_{\mathcal{H}}}  \cdots \frac{\tr (O_{u_q})}{D_{\mathcal{H}}}  \le \|O_0\| \prod_{s=1}^m \|O_s\|,
\end{align}
we obtain the upper bound 
\begin{align}
\copyave{m+1}{O_{0}^{(0)}  O_{1}^{(1)} O_{2}^{(2)}\cdots O_{m}^{(m)}} \le \mathcal{N}\left ( O_{0}^{(0)}  O_{1}^{(1)} O_{2}^{(2)}\cdots O_{m}^{(m)}  \right)
\|O_0\| \prod_{s=1}^m \|O_s\|.
\label{upp_O_product_extended_Hilbert_sp}
\end{align}

In order to estimate the upper bound of $\mathcal{N}\left ( O_{0}^{(0)}  O_{1}^{(1)} O_{2}^{(2)}\cdots O_{m}^{(m)}  \right)$, we consider a more general form as follows:
\begin{align}
\frac{1}{D_{\mathcal{H}}^{l_m+1}} \tr_{\mathcal{H}^{\otimes l_m+1}} \left ( O_{0}^{(0)}  O_{1}^{(l_1)} O_{2}^{(l_2)}\cdots O_{m}^{(l_m)}  \right)
=\copyave{l_m+1}{O_{0}^{(0)}  O_{1}^{(l_1)} O_{2}^{(l_2)}\cdots O_{m}^{(l_m)}} 
 ,
\end{align}
where $1\le l_1 < l_2 < \cdots < l_m < \infty$. 
We then aim to prove
\begin{align}
\mathcal{N}\left (O_{0}^{(0)}  O_{1}^{(l_1)} O_{2}^{(l_2)}\cdots O_{m}^{(l_m)}  \right) \le \prod_{s=1}^m 2N_{\ed_s |w_L} .\label{main_ineq_number_survive}
\end{align}
By applying the above inequality with $\{l_1,l_2,\ldots,l_m\}=\{1,2,\ldots,m\}$ to Ineq.~\eqref{upp_O_product_extended_Hilbert_sp}, we obtain the inequality~\eqref{ineq_Der_norm_overlap_form}

In the following, we give the proof of \eqref{main_ineq_number_survive} by using mathematical induction.
For $m=1$,  we have
\begin{align}
\copyave{l_1+1}{O_{0}^{(0)}  O_{1}^{(l_1)} } 
&= \tr_{\mathcal{H}^{\otimes l_1+1}} \left [ O_{0,\mathcal{H}_1} \left(-l_1 O_{1,\mathcal{H}_{l_1+1}}+ \sum_{j=1}^{l_1} O_{1,\mathcal{H}_j}\right) \right]  \notag \\
&= \tr (O_0 O_1) -  \tr (O_0) \cdot \tr( O_1)=\tr (O_0 O_1)   ,
\end{align}
which gives $\mathcal{N} ( O_{0}^{(0)}  O_{1}^{(1)}) =1$ as long as $\ed_0\cap \ed_1\neq \emptyset$, 
where we define $O_{1,\mathcal{H}_j}$ as in Eq.~\eqref{extended_operator_prop}.
We thus prove the inequality~\eqref{main_ineq_number_survive} for $m=1$.

We then prove the case of $m=M$ by assuming the inequality~\eqref{ineq_Der_norm_overlap_form} for $m=M-1$.
For the purpose, we introduce $\mathcal{C}_{s,s'}$ as a operation which applies to $O_{u,\mathcal{H}_j}$ ($j=1,2,\ldots,m+1$) as follows:
\begin{align}
\mathcal{C}_{s,s'} O_{u,\mathcal{H}_j}:=
\begin{cases}
0  & \for       \{s,s'\} \subset u,\\
O_{u,\mathcal{H}_j},& \quad        {\rm otherwise}
\end{cases}
\label{Def_mathcal_C_m_u}
\end{align}
where the operator $\mathcal{C}_{s,s'}$ acts on the index subspace $u \subset \{1,2,\ldots,m\}$.
Roughly speaking, the operator $\mathcal{C}_{s,s'}$ prohibits the two operator $O_s$ and $O_{s'}$ to be in the same Hilbert space.
  
From the definition~\eqref{Def_mathcal_C_m_u}, we obtain
\begin{align}
\mathcal{N}\left (\mathcal{C}_{s,s'} O_{0}^{(0)}  O_{1}^{(l_1)} O_{2}^{(l_2)} \cdots O_{M}^{(l_M)}\right)\le \mathcal{N}\left (O_{0}^{(0)} O_{1}^{(l_1)} O_{2}^{(l_2)} \cdots O_{M}^{(l_M)}  \right)
\label{Number_is_smaller_by_C_s_s'}
\end{align}
and
\begin{align}
\mathcal{C}_{s,s'} O_{s}^{(l_s)} O_{s'}^{(l_{s'})} =\mathcal{C}_{s,s'} O_{s'}^{(l_{s'})} O_{s}^{(l_s)} \quad {\rm or} \quad \mathcal{C}_{s,s'} \left[ O_{s}^{(l_s)}, O_{s'}^{(l_{s'})}\right]=0  ,\label{commutation_O_m_O_sC_m}
\end{align}
where $[\cdot, \cdot]$ is the commutator.
Also, the definition~\eqref{Def_mathcal_C_m_u} implies
\begin{align}
 O_{s}^{(l_s)} O_{s'}^{(l_{s'})} = (O_{s} O_{s'})^{(l_s)} +\mathcal{C}_{s,s'} O_{s}^{(l_s)} O_{s'}^{(l_{s'})}   \label{decomp_two_op_C_m}
\end{align}
for $l_s < l_{s'}$. 

We here denote by $O_{s_0}$ the operator that has the minimum $N_{\ed_{s} |w_L}$, namely $N_{\ed_{s_0} |w_L} \le N_{\ed_{s} |w_L}$ ($s=0,1,2, \ldots, M$).
We also define $ \{s_1,s_2,\ldots,s_{\tilde{m}}\}$ ($s_1\le s_2\le \cdots \le s_{\tilde{m}}$) as the indices which satisfy $\ed_{s_j} \cap \ed_{s_0} \neq \emptyset$; notice that $\tilde{m}=N_{\ed_{s_0} |w_L}$.
In the following, we assume $s_0 \le s_1$ for the simplicity, but the same discussion is applied to the general cases. 
Because $O_{s_0}$ commutes with $O_s$ if $s\notin \{s_1,s_2,\ldots,s_{\tilde{m}}\}$, we have
\begin{align}
O_{0}^{(0)}  O_{1}^{(l_1)} O_{2}^{(l_2)} \cdots O_{M}^{(l_M)}  =& O_{0}^{(0)}  O_{1}^{(l_1)} \cdots  O_{s_1-1}^{(l_{s_1-1})}  O_{s_0}^{(l_{s_0})} O_{s_1}^{(l_{s_1})}  \cdots O_{M}^{(l_M)}  \notag \\
 =&O_{0}^{(0)}   O_{1}^{(l_1)} \cdots  \left(O_{s_0} O_{s_1} \right)^{(l_{s_0})} \cdots O_{M}^{(l_M)}  +\mathcal{C}_{s_0,s_1}   O_{0}^{(0)}  O_{1}^{(l_1)} O_{2}^{(l_2)} \cdots O_{M}^{(l_M)},
\end{align}
where in the second equality we use Eq.~\eqref{decomp_two_op_C_m}.
In the same way, from Eq.~\eqref{commutation_O_m_O_sC_m}, we have
 \begin{align}
\mathcal{C}_{s_0,s_1} O_{0}^{(0)}  O_{1}^{(l_1)} O_{2}^{(l_2)} \cdots O_{M}^{(l_M)} 
&=\mathcal{C}_{s_0,s_1} O_{0}^{(0)}  O_{1}^{(l_1)}  \cdots  \left(O_{s_0}O_{s_2} \right)^{(l_{s_0})} \cdots O_{M}^{(l_M)}  + \mathcal{C}_{s_0,s_1}\mathcal{C}_{s_0,s_2}     O_{0}^{(0)}  O_{1}^{(l_1)} O_{2}^{(l_2)} \cdots O_{M}^{(l_M)} .
\end{align}
By repeating this process, we finally obtain
\begin{align}
O_{0}^{(0)}  O_{1}^{(l_1)} O_{2}^{(l_2)} \cdots O_{M}^{(l_M)} &=O_{0}^{(0)}   O_{1}^{(l_1)} \cdots  \left(O_{s_0} O_{s_1} \right)^{(l_{s_0})} \cdots O_{M}^{(l_M)}   \notag \\
 &+\mathcal{C}_{s_0,s_1} O_{0}^{(0)}  O_{1}^{(l_1)}  \cdots  \left(O_{s_0}O_{s_2} \right)^{(l_{s_0})} \cdots O_{M}^{(l_M)}  \notag \\
 &+  \mathcal{C}_{s_0,s_1}\mathcal{C}_{s_0,s_2}  O_{0}^{(0)}  O_{1}^{(l_1)}  \cdots  \left(O_{s_0}O_{s_3} \right)^{(l_{s_0})} \cdots O_{M}^{(l_M)}  \notag \\
 &+\cdots + \mathcal{C}_{s_0,s_1}\mathcal{C}_{s_0,s_2} \cdots  \mathcal{C}_{s_0,s_{\tilde{m}-1}}  O_{0}^{(0)}  O_{1}^{(l_1)}  \cdots  \left(O_{s_0}O_{s_{\tilde{m}}} \right)^{(l_{s_0})} \cdots O_{M}^{(l_M)}   \notag \\
 &+ \mathcal{C}_{s_0,s_1}\mathcal{C}_{s_0,s_2} \cdots  \mathcal{C}_{s_0,s_{\tilde{m}}} O_{0}^{(0)}  O_{1}^{(l_1)} O_{2}^{(l_2)} \cdots O_{M}^{(l_M)}. 
\label{decomp_O_product_C}
\end{align}
From $\tr(O_s) =0$ for $s=0,1,\ldots, M$, we obtain
$
 \tr \left(  \mathcal{C}_{s_0,s_1}\mathcal{C}_{s_0,s_2} \cdots  \mathcal{C}_{s_0,s_{\tilde{m}}} O_{0}^{(0)}  O_{1}^{(l_1)} O_{2}^{(l_2)} \cdots O_{M}^{(l_M)}\right)=0,
$
and hence 
\begin{align}
\mathcal{N}\left(  \mathcal{C}_{s_0,s_1}\mathcal{C}_{s_0,s_2} \cdots  \mathcal{C}_{s_0,s_{\tilde{m}}} O_{0}^{(0)}  O_{1}^{(l_1)} O_{2}^{(l_2)} \cdots O_{M}^{(l_M)}\right)=0.
\label{N_C_m_vanish}
\end{align}
By combining the equation~\eqref{decomp_O_product_C} with the relations~\eqref{Number_is_smaller_by_C_s_s'} and \eqref{N_C_m_vanish}, we have
\begin{align}
  \mathcal{N} \left( O_{0}^{(0)}  O_{1}^{(l_1)} O_{2}^{(l_2)} \cdots O_{M}^{(l_M)}  \right) &\le 
 \mathcal{N} \left(  O_{0}^{(0)}   O_{1}^{(l_1)} \cdots  \left(O_{s_0} O_{s_1} \right)^{(l_{s_0})} \cdots O_{M}^{(l_M)} \right) \notag \\
 &+ \mathcal{N} \left(O_{0}^{(0)}  O_{1}^{(l_1)}  \cdots  \left(O_{s_0}O_{s_2} \right)^{(l_{s_0})} \cdots O_{M}^{(l_M)} \right)  \notag \\
 &+  \mathcal{N} \left( O_{0}^{(0)}  O_{1}^{(l_1)}  \cdots  \left(O_{s_0}O_{s_3} \right)^{(l_{s_0})} \cdots O_{M}^{(l_M)}  \right) \notag \\
 &+\cdots +  \mathcal{N} \left(  O_{0}^{(0)}  O_{1}^{(l_1)}  \cdots  \left(O_{s_0}O_{s_{\tilde{m}}} \right)^{(l_{s_0})} \cdots O_{M}^{(l_M)} \right)  . \label{m-1_sum_s_0_N_s_0}
\end{align}

We now upperbound each of the terms in the right-hand side of \eqref{m-1_sum_s_0_N_s_0}.
The assumption of Ineq.~\eqref{main_ineq_number_survive} for $m=M-1$ implies the upper bound of
\begin{align}
 \mathcal{N} \left(  O_{0}^{(0)}   O_{1}^{(l_1)} \cdots  \left(O_{s_0} O_{s_1} \right)^{(l_{s_0})} \cdots O_{M}^{(l_M)} \right)
 &\le  2N_{\ed_{s_0} \cup \ed_{s_1} | w_L} \prod_{s\neq s_0,s_1} 2N_{\ed_s | w_L}\notag \\
 & \le2 (N_{\ed_{s_0}|w_L} + N_{\ed_{s_1}|w_L}) \prod_{s\neq s_0,s_1} 2N_{\ed_s | w_L} \le 2 \prod_{s\neq s_0} 2N_{\ed_s | w_L} ,\label{m-1_sum_s_0_s_1}
\end{align}
where we use $N_{\ed_{s_0} \cup \ed_{s_1}|w_L} \le N_{\ed_{s_0}|w_L} + N_{\ed_{s_1}|w_L} \le2N_{\ed_{s_1}|w_L} $ due to $N_{\ed_{s_0} |w_L} \le N_{\ed_{s} |w_L}$ ($s=0,1,2, \ldots, M$).
By combining the two inequalities~\eqref{m-1_sum_s_0_N_s_0} and \eqref{m-1_sum_s_0_s_1}, we finally obtain 
\begin{align}
  \mathcal{N} \left( O_{0}^{(0)}  O_{1}^{(l_1)} O_{2}^{(l_2)} \cdots O_{M}^{(l_M)}  \right) &\le  2 \tilde{m}  \prod_{s\neq s_0} 2N_{\ed_s|w_L} = \prod_{s=1}^M 2N_{\ed_s|w_L} .
\end{align} 
This completes the proof of the inequality~\eqref{main_ineq_number_survive}.
 $\square$


%
%

\section {Proof of Proposition~\ref{prop: convergence of cluster expansion}} \label{proof_prop: convergence of cluster expansion}

\begin{figure}
\centering
\includegraphics[clip, scale=0.6]{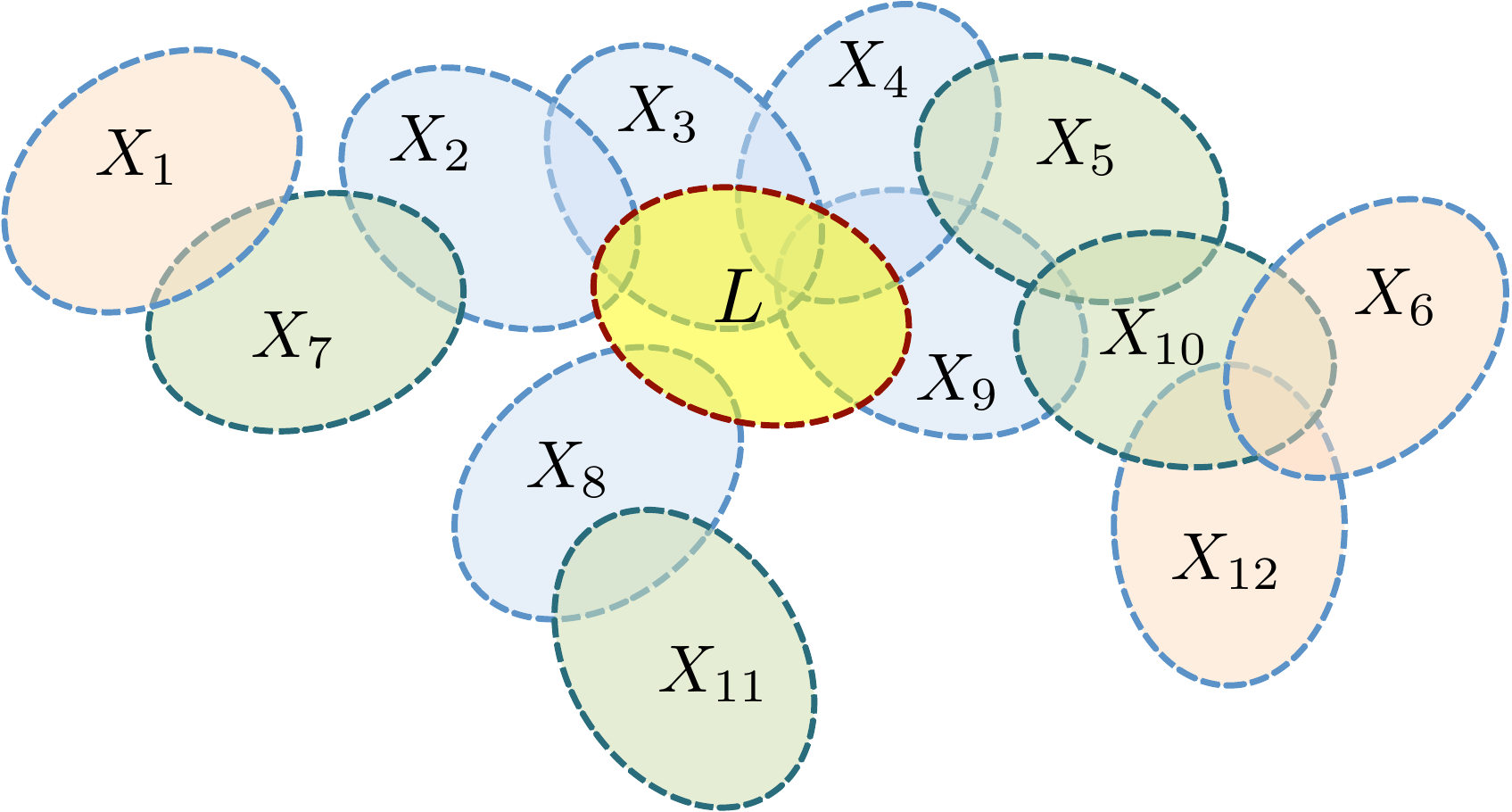}
\caption{Decomposition of $w_L$ in Eq.~\eqref{w_L_decomposition_cal_cluster_sum}.
In the picture, we have $w_0=\{L\}$, $w_1=\{\ed_2,\ed_3,\ed_4,\ed_8,\ed_9\}$, $w_2=\{\ed_5,\ed_7,\ed_{10},\ed_{11}\}$ and $w_3=\{\ed_1,\ed_6,\ed_{12}\}$.
}
\label{fig:connection_hierarchy}
\end{figure}

We here obtain an upper bound of 
\begin{align}
\sum_{w \in \Gc^L_{m}} \frac{n_{w}}{m!} \prod_{s=1}^m N_{\ed_s | w_L} \| o_{\ed_s}\|. \label{start_inequality_for_conv_thm}
 \end{align}
In order to estimate the summation, we first decompose $w_L$ as follows:
\begin{align}
w_L= w_0 \oplus w_1 \oplus w_2 \oplus \cdots \oplus w_l, \quad 1\le l\le m,\label{w_L_decomposition_cal_cluster_sum}
\end{align}
where $w_0=\{L\}$ and $w_j \subset w_L $ satisfy $\dist(w_j, w_0) = j$ for $j=1,2,\ldots,l$. 
Here, we define $\dist(w_j, w_0)$ as the shortest path length in the cluster $w_0\oplus w_1 \oplus \cdots \oplus w_{j-1}$ which connects from $w_j$ to $w_0$. 
We also define $q_j:=|w_j|$ with $q_j\ge 1$.
We then obtain
\begin{align}
  \sum_{w\in \Gc_m^L }n_w  =\sum_{l=1}^m \sum_{\{q_1,q_2,\ldots,q_l\}}\frac{n_w}{n_{w_1}n_{w_2} \cdots n_{w_l}}
 \sum_{w_{1}|w_{0}}n_{w_1}  \sum_{w_{2}|w_0,w_{1}}n_{w_2} \cdots   \sum_{w_l| w_0,w_1, \ldots, w_{l-1}} n_{w_l} , \label{w_L_connect_express}
\end{align}
where $\sum_{w_{j}|w_0,w_1, \ldots, w_{j-1}}$ denotes the summation with respect to $w_j$ such that $\dist(w_j, w_0) = j$ when $\{w_1,w_2,\ldots, w_{j-1}\}$ are fixed.
Note that $w_0$ has been already fixed as $w_0=\{L\}$.
Also, because the subsets in $w_j$ have overlaps with those only in $w_{j-1}$, $w_j$ and $w_{j+1}$, we have
\begin{align}
N_{\ed^{(j)}_s|w_L} =  N_{\ed^{(j)}_s|w_{j-1} \oplus w_{j}\oplus w_{j+1} }   \for s =1,2 ,\ldots q_j,
\end{align}
where we denote $w_j=\{\ed^{(j)}_s\}_{s=1}^{q_j}$.

By using the above decomposition, we can reduce the inequality~\eqref{start_inequality_for_conv_thm} to
\begin{align} 
\sum_{w \in \Gc^L_{m}} \frac{n_{w}}{m!} &\prod_{s=1}^m N_{\ed_s | w_L} \| o_{\ed_s}\|\le  \frac{1}{m!}\sum_{l=1}^m \sum_{\{q_1,q_2,\ldots,q_l\}}\frac{m!}{q_1! q_2! \dots q_l!}
 \sum_{w_{1}|w_{0}}n_{w_1} \prod_{s_1=1}^{q_1}  N_{\ed^{(1)}_{s_1}|w_{0} \oplus w_{1}\oplus w_{2} }     \left\| o_{\ed^{(1)}_{s_1}}\right \|     \times  \notag \\
 &    \sum_{w_{2}|w_0,w_{1}}n_{w_2 }\prod_{s_2=1}^{q_2}   N_{\ed^{(2)}_{s_2}|w_{1} \oplus w_{2}\oplus w_{3} }  \left\| o_{\ed^{(2)}_{s_2}}\right\| 
 \times \cdots  \times \sum_{w_l|w_0, \ldots, w_{l-1}} n_{w_l}\prod_{s_l=1}^{q_l}   N_{\ed^{(l)}_{s_l}|w_{l-1} \oplus w_{l} } \left\| o_{\ed^{(l)}_{s_l}}\right\|,
 \label{upper_bound_sum_w_j_all}
\end{align}
where we used the equation of
 \begin{align} 
\frac{n_w}{n_{w_1}n_{w_2} \cdots n_{w_l}} = \frac{m!}{q_1! q_2! \dots q_l!},
 \end{align} 
which is derived from the fact that the elements of $w_i$ and $w_j$ are different for arbitrary $i,j \in \{1,2,\ldots, l\}$. 
 
In the following, we aim to calculate the upper bound of 
\begin{align} 
\sum_{w_j|w_0, \ldots,w_{j-1}} n_{w_j} \prod_{s_j=1}^{q_j}   N_{\ed^{(j)}_{s_j}|w_{j-1} \oplus w_{j}\oplus w_{j+1} } \left\| o_{\ed^{(j)}_{s_j}}\right\|
\end{align}
only by using $q_{j-1},q_j,q_{j+1}$, which does not depend on the details of $w_{j-1},w_j,w_{j+1}$.
For the purpose, we start from the summation with respect to $w_l$:
\begin{align} 
& \sum_{w_{l-1}|w_0, \ldots, w_{l-2}}n_{w_{l-1}} \prod_{s_{l-1}=1}^{q_{l-1}}   N_{\ed^{(l-1)}_{s_{l-1}}| w_{l-2} \oplus w_{l-1}\oplus w_{l} } \left\| o_{\ed^{(l-1)}_{s_{l-1}}}\right\|    
 \sum_{w_l|w_0, \ldots, w_{l-1}}n_{w_l} \prod_{s_l=1}^{q_l}   N_{\ed^{(l)}_{s_l}|w_{l-1} \oplus w_{l} } \left\| o_{\ed^{(l)}_{s_l}}\right\| \notag \\
 \le&  \left(\max_{\tilde{w}_l\in \Cl_{q_l}}  \sum_{w_{l-1}|w_0, \ldots, w_{l-2}}n_{w_{l-1}} \prod_{s_{l-1}=1}^{q_{l-1}}  N_{\ed^{(l-1)}_{s_{l-1}}| w_{l-2} \oplus w_{l-1}\oplus \tilde{w}_l } \left\| o_{\ed^{(l-1)}_{s_{l-1}}}\right\|    \right) \notag \\
&\times\left(   \sum_{w_l|w_0, \ldots, w_{l-1}}n_{w_l} \prod_{s_l=1}^{q_l}  N_{\ed^{(l)}_{s_l}|w_{l-1} \oplus w_{l} } \left\| o_{\ed^{(l)}_{s_l}}\right\|  \right).
\end{align}
For fixed $\{w_1,w_2,\ldots, w_{l-1}\}$, we aim to obtain an upper bound of 
\begin{align} 
 \sum_{w_l|w_0, \ldots,w_{l-1}}n_{w_l} \prod_{s_l=1}^{q_l}  N_{\ed^{(l)}_{s_l}|w_{l-1} \oplus w_{l} } \left\| o_{\ed^{(l)}_{s_l}}\right\|  ,
\end{align}
only by using $q_l$ and $q_{l-1}$.

First, we have
\begin{align} 
 \sum_{w_l|w_0, \ldots,w_{l-1}}n_{w_l} \prod_{s_l=1}^{q_l}  N_{\ed^{(l)}_{s_l}|w_{l-1} \oplus w_{l} }\left\| o_{\ed^{(l)}_{s_l}}\right\| 
 &\le  \prod_{s_l=1}^{q_l} \left(\sum_{\ed^{(l)}_{s_l}: \ed^{(l)}_{s_l} \cap V_{w_{l-1}}\neq 0}  N_{\ed^{(l)}_{s_l}|w_{l-1} \oplus w_{l} }\left\| o_{\ed^{(l)}_{s_l}}\right\| \right)
\notag \\
 & \le \left(\max_{\tilde{w}_l\in \Cl_{q_l}} \sum_{\ed: \ed \cap V_{w_{l-1}} \neq 0}  N_{\ed|w_{l-1} \oplus \tilde{w}_{l} } \left\| o_{\ed}\right\|  \right)^{q_l}.
 \label{Ineq: w_l_summation_1}
\end{align}
We then prove the following inequality for arbitrary $w \in \Cl_{|w|}$:
\begin{align} 
\sum_{\ed: \ed \cap V_{w} \neq \emptyset  } N_{\ed|w} \left\| o_{\ed}\right\|  \le g  \sum_{j=1}^{|w|} |\ed_j| \for \forall w , \label{sum0_overlap_cal}
\end{align}
where we denote $w=\{\ed_j\}_{j=1}^{|w|}$.
For the proof of \eqref{sum0_overlap_cal}, we start from the inequality as
\begin{align} 
\sum_{\ed: \ed \cap V_{w} \neq \emptyset} N_{\ed|w} \left\| o_{\ed}\right\|  \le \sum_{v \in V_{w}} \sum_{\ed:\ed\ni v} n_{v|w}\|o_\ed\| \label{sum1_overlap_cal}
\end{align}
where $n_{v|w}$ is a number of subsets in $w$ which have overlaps with the spin $v$.  
From the condition~\eqref{cond_for_o_ed}, we obtain
 \begin{align} 
\sum_{v \in V_{w}} \sum_{\ed:\ed\ni v} n_{v|w}\|o_\ed\| \le  g \sum_{v\in V_{w}} n_{v|w} =g \sum_{j=1}^{|w|} |\ed_j|. \label{sum2_overlap_cal}
\end{align}
Finally, by combining the inequalities~\eqref{sum1_overlap_cal} and \eqref{sum2_overlap_cal}, we obtain the inequality~\eqref{sum0_overlap_cal}.
By using the inequality~\eqref{sum0_overlap_cal} with $w=w_{l-1} \oplus \tilde{w}_{l}$, the inequality~\eqref{Ineq: w_l_summation_1} reduces to
\begin{align} 
 &\sum_{w_l|w_{l-1}} \prod_{s_l=1}^{q_l}  N_{\ed^{(l)}_{s_l}|w_{l-1} \oplus w_{l} }\left\| o_{\ed^{(l)}_{s_l}}\right\|  \le \left[ g k ( q_{l-1}+q_{l}) \right]^{q_l},\label{Ineq: w_l_summation_1}
\end{align}
where we use the fact that the cardinality of $\ed \in E_k$ satisfies $|\ed|\le k$.

After the summation with respect to $w_l$, we can apply the same calculation for the summation with respect to $w_{l-1}$ for a fixed $w_{l-2}$:
\begin{align} 
 &  \sum_{w_{l-2}|w_0, \ldots,w_{l-3}}n_{w_{l-2}}\prod_{s_{l-2}=1}^{q_{l-2}}  N_{\ed^{(l-2)}_{s_{l-2}}| w_{l-3} \oplus w_{l-2}\oplus w_{l-1} } \left\| o_{\ed^{(l-2)}_{s_{l-2}}}\right\|  \notag \\
 &\times \left(\max_{\tilde{w}_l\in \Cl_{q_l}}  \sum_{w_{l-1}|w_0, \ldots,w_{l-2}}n_{w_{l-1}} \prod_{s_{l-1}=1}^{q_{l-1}}  N_{\ed^{(l-1)}_{s_{l-1}}| w_{l-2} \oplus w_{l-1}\oplus \tilde{w}_l } \left\| o_{\ed^{(l-1)}_{s_{l-1}}}\right\|    \right)  \notag \\
&\le \left(\max_{\tilde{w}_{l-1}\in \Cl_{q_{l-1}}}  \sum_{w_{l-2}|w_0, \ldots,w_{l-3}}n_{w_{l-2}} \prod_{s_{l-2}=1}^{q_{l-2}}  N_{\ed^{(l-2)}_{s_{l-2}}| w_{l-3} \oplus w_{l-2}\oplus \tilde{w}_{l-1} } \left\| o_{\ed^{(l-2)}_{s_{l-2}}}\right\|   \right)  \notag \\
&\times \left(\max_{\tilde{w}_l\in \Cl_{q_l}}  \sum_{w_{l-1}|w_0, \ldots,w_{l-2}} n_{w_{l-1}} \prod_{s_{l-1}=1}^{q_{l-1}}  N_{\ed^{(l-1)}_{s_{l-1}}| w_{l-2} \oplus w_{l-1}\oplus \tilde{w}_l } \left\| o_{\ed^{(l-1)}_{s_{l-1}}}\right\|    \right). 
\end{align}
In the same way as the derivation of Ineq.~\eqref{Ineq: w_l_summation_1}, we obtain
\begin{align} 
 &\max_{\tilde{w}_l\in \Cl_{q_l}}  \sum_{w_{l-1}|w_0, \ldots,w_{l-2}}n_{w_{l-1}} \prod_{s_{l-1}=1}^{q_{l-1}}  N_{\ed^{(l-1)}_{s_{l-1}}| w_{l-2} \oplus w_{l-1}\oplus \tilde{w}_l } \left\| o_{\ed^{(l-1)}_{s_{l-1}}}\right\|   \le \left[ g k (q_{l-2}+ q_{l-1}+q_{l}) \right]^{q_{l-1}}.\label{Ineq: w_l_summation_12}
\end{align}

By repeating this process, we finally obtain
\begin{align} 
\sum_{w \in \Gc^L_{m}} \frac{n_{w}}{m!} \prod_{s=1}^m N_{\ed_s | w_L} \left\| o_{\ed_s} \right\|  
\le\sum_{l=1}^m \sum_{\{q_1,q_2,\ldots,q_l\}}\frac{1 }{q_1! q_2! \dots q_l!}
\left[g(|L|+kq_1+ kq_{2})  \right]^{q_1}   \prod_{j=2}^{l}  \left[gk(q_{j-1}+q_j+ q_{j+1})  \right]^{q_j}  ,
 \label{upper_bound_sum_w_j_all}
\end{align}
where we set $q_{l+1}=0$. By using $n! \ge (n/e)^n$ and $e^{q_j} [(q_{j-1}+q_j+q_{j+1})/q_j]^{q_j} \le e^{q_{j-1}+q_j+q_{j+1}}$, we have
\begin{align} 
&\frac{1}{q_1! q_2! \dots q_l!}\left[g(|L|+kq_1+ kq_{2})  \right]^{q_1} \prod_{j=2}^{l}  \left[gk(q_{j-1}+q_j+ q_{j+1})  \right]^{q_j}  \notag \\
\le& (gk)^m e^{|L|/k}\exp\left(\sum_{j=1}^l (q_{j-1}+q_j+q_{j+1}) \right)\le e^{|L|/k} (e^{3} gk)^m,
\end{align}
which yields 
\begin{align} 
\sum_{w \in \Gc^L_{m}} \frac{n_{w}}{m!} \prod_{s=1}^m N_{\ed_s | w_L} \left\| o_{\ed_s} \right\|    \le& \sum_{l=1}^m \sum_{\{q_1,q_2,\ldots,q_l\}}e^{|L|/k} (e^{3} gk)^m.
\end{align}
The summation with respect to $\{q_1,q_2,\ldots,q_l\}$ is equal to the $(m-l)$-multicombination from a set of $l$ elements: 
\begin{align} 
\sum_{\{q_1,q_2,\ldots,q_l\}}=\multiset{l}{m-l} = \binom{m-1}{l-1} \label{comb_q_1_q_2__q_l},
\end{align}
and hence 
\begin{align} 
\sum_{w \in \Gc^L_{m}} \frac{n_{w}}{m!} \prod_{s=1}^m N_{\ed_s | w_L} \left\| o_{\ed_s} \right\|  \le& \sum_{l=1}^m \binom{m-1}{l-1} e^{|L|/k} (e^{3} gk)^m 
 =\frac{1}{2}  e^{|L|/k} (2e^{3} gk)^m.
\end{align}
This completes the proof of Proposition~\ref{prop: convergence of cluster expansion}. $\square$

%
%


\bibliography{CH_long.bib}

\end{document}